\documentclass[12pt]{article}   %%LaTeX2e

\usepackage{psfig,cite}
\newcommand{\dms}{\displaystyle}  
\newenvironment{comment}[1]{}{}

\newcommand{\equ}[1]{Eq.\,(\ref{#1})}
\newcommand{\eqs}[1]{Eqs.\,(\ref{#1})}

\newcommand{\ew}{electroweak~}
\newcommand{\non}{\nonumber}
\newcommand{\gsim}{\;\rlap{\lower 3.5 pt \hbox{$\mathchar \sim$}} \raise 1pt
 \hbox {$>$}\;}
\newcommand{\lsim}{\;\rlap{\lower 3.5 pt \hbox{$\mathchar \sim$}} \raise 1pt
 \hbox {$<$}\;}
\newcommand{\smallz}{{\scriptscriptstyle Z}} %  a smaller Z
\newcommand{\smalla}{{\scriptscriptstyle A}} 
\newcommand{\smallw}{{\scriptscriptstyle W}} %
\newcommand{\smallh}{{\scriptscriptstyle H}} %

\newcommand{\smallT}{{\scriptscriptstyle T}}
\newcommand{\smallS}{{\scriptscriptstyle S}}
\newcommand{\smallL}{{\scriptscriptstyle L}}
\newcommand{\smallR}{{\scriptscriptstyle R}}
\newcommand{\smallD}{{\scriptscriptstyle D}}
\newcommand{\smallB}{{\scriptscriptstyle B}}
\newcommand{\smallG}{{\scriptscriptstyle G}}
\newcommand{\mz}{M_\smallz}
\newcommand{\mw}{M_\smallw}
\newcommand{\mh}{M_\smallh}
\newcommand{\mt}{M_t}
\def\pl#1#2#3{{\it Phys. Lett. }{\bf B#1~}(19#2)~#3}
\def\zp#1#2#3{{\it Z. Phys. }{\bf C#1~}(19#2)~#3}
\def\prl#1#2#3{{\it Phys. Rev. Lett. }{\bf #1~}(19#2)~#3}

\def\pr#1#2#3{{\em Phys. Rev. }{\bf D#1~}(19#2)~#3}
\def\np#1#2#3{{\em Nucl. Phys. }{\bf B#1~}(19#2)~#3}
\def\ap#1#2#3{{\it Ann. Phys.} (NY) #1 (19#2) #3}
\def\cmp#1#2#3{{\it Comm. Math. Phys.} #1 (19#2) #3}

\newcommand{\efe}[1]{Ref.\cite{#1}}

\newcommand{\be}{\begin{equation}}
\newcommand{\ee}{\end{equation}}
\newcommand{\een}{\end{subequations}}
\newcommand{\ben}{\begin{subequations}}
\newcommand{\beq}{\begin{eqalignno}}
\newcommand{\eeq}{\end{eqalignno}}
\newcommand{\bea}{\begin{eqnarray}}
\newcommand{\eea}{\end{eqnarray}}
\newcommand{\ber}{\begin{array}} \newcommand{\eer}{\end{array}}

\newcommand{\dx}{d^{4}x}

          \newcommand{\gam}{\gamma}    

                \newcommand{\del}{\delta}    
\def\wt{\widetilde}
 \newcommand{\g}{{\bf  \Gamma}}                  
 \newcommand{\gh}{\hat{\g}}

\def\b#1{\overline{#1}}  
  
  \newenvironment{appendletterA}
 {
  \typeout{ Starting Appendix \thesection }
  \setcounter{equation}{0}
  
}{
  \typeout{Appendix done}
 }
  \newenvironment{appendletterB}
 {
  \typeout{ Starting Appendix \thesection }
  \setcounter{equation}{0}
  
}{
  \typeout{Appendix done}
 }
  \newenvironment{appendletterC}
 {
  \typeout{ Starting Appendix \thesection }
  \setcounter{equation}{0}
  
}{
  \typeout{Appendix done}
 }

\def\bc{\bar{c}}

\newcommand{\msbar}{\overline{\rm MS}}
\def \non {\nonumber}
\def \mt   {M_t}

\def \ms   {\overline{\mbox{MS}}}

\newcommand{\de}{\partial}  

\begin{document}              
\begin{titlepage}
\begin{flushright}
        \small
        TUM-HEP-352/99\\
        MPI-PhT/99-24\\
        June 1999
\end{flushright}

\begin{center}
\vspace{1cm}
{\Large\bf The Nielsen Identities of the SM \\ and  the definition of mass} 

\vspace{0.5cm}
\renewcommand{\thefootnote}{\fnsymbol{footnote}}
{\bf           P.~Gambino$^a$ and 
                        P.A.~Grassi$^b$}
\setcounter{footnote}{0}
\vspace{.8cm}

{\it
        $^a$ Technische Universit{\"a}t M{\"u}nchen,\\
                       Physik Dept., D-85748 Garching, Germany \\
\vspace{2mm}
        $^b$ Max Planck Institut f{\"u}r Physik (Werner-Heisenberg-Institut),\\
        F{\"o}hringer Ring 6, D-80805 Munich,  Germany }
\vspace{1.5cm}

{\large\bf Abstract}
\end{center}
In a generic gauge theory the gauge parameter dependence of individual Green
functions is controlled by the Nielsen identities, which originate from an
enlarged BRST symmetry. We give a practical introduction to the Nielsen
identities of the Standard Model (SM) and to their renormalization
and illustrate the power of this elegant 
formalism in the case of the problem of the definition of mass.
We prove to all orders in perturbation theory the gauge-independence of the
complex pole of the propagator 
for all physical fields of the SM, in the most general case with
mixing and CP violation. At the amplitude level, the formalism provides an
intuitive and  general understanding of the gauge recombinations
which makes it particularly useful at higher orders.
We also include in an appendix the explicit expressions 
for  the fermionic two-point functions in a generic $R_\xi$ gauge.
\noindent

% PACS numbers: 

\end{titlepage}
\newpage
\section{Introduction}
\label{intro}
Considering  the subtle cancellations between various contributions 
necessary  to make physical observables gauge-parameter independent,
it is  not surprising that the variation of individual  Green
functions with respect to the gauge-fixing parameters
are governed by symmetry relations.
 Formally,  these relations can be shown to 
follow from an enlarged BRST symmetry in which the gauge
parameters also undergo a BRST transformation \cite{zuber,si-pig}.
They are non-linear identities of the same kind of the Slavnov-Taylor 
Identities
(STI), satisfied by  Green functions at arbitrary  external momenta,
and are usually called Nielsen identities, after the seminal  paper
\cite{nielsen} in which they were first presented.

The power of this technique lies in the possibility of factorizing
the gauge parameter dependence in terms of  new 
objects, the Green functions of  BRST sources associated to the gauge
parameters. 
In principle, this factorization holds at any order in perturbation theory, 
but its interplay with the renormalization procedure is not trivial.
In the case of gauge-independent quantities, the gauge 
cancellations emerge from the recombination between these new objects 
and can be  verified   without  an explicit evaluation of multi-loop diagrams.
As we will  see in the following, 
the mechanism of gauge recombination is revealed in great simplicity 
in the case of physical amplitudes.

The Nielsen identities provide the appropriate framework to study any problem
related to gauge dependence. They 
are well known to field theory experts and have been 
 used in the study of the effective potential 
\cite{nielsen,aitchison} and in high temperature field theory \cite{HT}.
 Recently, they have also been studied in the context of 
the Abelian Higgs model \cite{kraus-si} and of  Yang-Mills theories 
\cite{kraus-si-h} with background fields. 
Our main purpose in this paper  is to introduce
the Nielsen identities of the full Standard Model (SM) 
and to study their renormalization.
From a conceptual point of view,  if all the physical parameters are fixed 
by  normalization conditions  {\it directly} based on physical observables,
using the Nielsen identities it is possible to  make sure 
that other quantities are gauge-independent. 
This applies to mass and mixing parameters of unstable fields
--- which we consider in detail --- as well as to
off-shell  objects like  effective charges.
From a more practical point of view, we believe the Nielsen 
identities are also a useful tool for multi-loop calculations both in
the electroweak SM  and in QCD. The identities for the two, three, and
four point functions that we obtain in the present paper can also be
useful in this context.
Throughout the paper, we will proceed in a pedagogical way and complement the 
formal treatment with explicit one-loop examples.

As a demonstrative ground for the technique of the Nielsen identities we 
have chosen  the problem of the definition of mass in the SM. This is an
important and non-trivial issue 
which recently has received renewed attention 
\cite{zmass,hveltman,psw,kronfeld},  prompted in part by the 
high precision measurements of the $Z^0$ mass at LEP and SLC.
It has been shown long ago \cite{veltman} that unstable particles
are compatible with unitarity and causality. 
However, what makes the perturbative definition of the parameters
associated to unstable fields
a delicate and intriguing problem is the interplay between 
the phenomenon of resonance (which goes beyond perturbation theory
as it implies the Dyson summation of an infinite number of diagrams)
and the perturbative implementation of gauge symmetry.
In particular, the correct identification of the mass parameters 
of an unstable particle is not 
obvious: their gauge independence must be proved in full generality
and their connection to experimental quantities clarified.  

A concept which is generally believed to be gauge-independent is the complex
pole of the resummed propagator. To the best of our knowledge, 
there exists  no general and rigorous proof that this is true.
In this paper we use  the Nielsen identities to   provide the proof
to all orders in perturbation theory and for all physical fields of the SM. 
The only  assumption clearly needed to obtain  this result 
is that the renormalization conditions for the {\it physical}
parameters do not introduce spurious gauge dependence. This is the case 
whenever they are based on a well-defined set of physical observables.
We also discuss  how a mass parameter for the unstable fields
can be consistently defined on this basis.

We have organized the paper in the following way. In the next section we 
introduce the Nielsen Identity for the 1PI generating functional at the 
classical level. After a discussion of the renormalization conditions,  
we study  the modifications introduced by quantum corrections
in the most general scenario. In Sec.\,3, as an elementary illustration,
we discuss the Nielsen identities for the one-point 
Green functions. %
In Sec.\,4 we consider the case of the $W$ boson and 
prove the gauge-parameter independence of the complex pole of its propagator.
Several comments and examples here should help clarify the most important points.
As a digression, we also consider the infrared finiteness of the $W$ pole mass.
The analysis is then extended to the case of mixing. 
In Sec.\ref{gammaZ} we consider the $\gamma,Z^0$ sector and derive an
interesting relation for the photon correlator at $q^2=0$ in the SM.
We  then study  
in Sec.\,\ref{scalar}  the scalar sector and in Sec.\,\ref{fermions}
 the fermionic sector. The following section is devoted to a discussion of
the mechanism of gauge-cancellations in the case of four-fermion processes.
Sec.\,\ref{summary} 
concludes the main text summarizing the most important points.
We have collected some useful material in three appendices: 
in the first one we 
discuss some aspects of the derivation of the Nielsen identities and
present the sector of the Lagrangian containing the BRST sources.
In App.\,B we give a technical proof that is crucial for the 
results of Sec.\,2.
Finally, we provide in App.\,C the full one-loop fermionic self-energies in an
arbitrary $R_\xi$ gauge. This completes the work of \efe{ds92}, where the 
one-loop gauge dependence of the basic electroweak corrections 
has been considered.

\section{The Nielsen identities in the SM}
\label{sec2}
The idea behind the Nielsen identities is  simple: the variation of 
the classical action with respect to a gauge parameter coincides with the BRST 
variation of a local polynomial in the fields. This is clearly necessary 
in order to guarantee the gauge-independence of physical observables.
For example, the variation of an S-matrix  element with respect 
to the gauge parameters corresponds to the  insertion of the 
BRST variation of a 
local  term between physical states,  which is known to vanish. 
The Nielsen identities  implement this simple idea at the quantum level.  

Our starting point is the Nielsen identity for the 
generating functional $\g$ at the classical level \cite{si-pig,nielsen}, 
\be
\frac{\de}{\de\xi}
\g_0 =  {{\cal S}_{\Gamma_0}} \left( \frac{\de}{\de \chi}\,\g_0   \right)
\label{nielsen},
\ee  
where $\chi=s\,\xi$ is the BRST source associated 
to a  generic gauge parameter
$\xi$, $ s$ is the classical BRST
generator, and ${\cal S}_{\Gamma_0}$  the linearized Slavnov-Taylor operator whose definition is recalled  in  Appendix A.
Notice that the operator coupled to $\chi$ is non-linear in the quantum fields, therefore it 
requires a proper renormalization.  The extended BRST 
automatically takes into account the renormalization of the theory and the renormalization of the 
composite operators generated by the variations of the action with respect to the gauge 
parameters. Here and henceforth, we used the reduced functional, also defined in App. A,
in place  of the standard  functional of proper functions. 
In the case of linear gauges, this allows us to 
write STI and Nielsen identities in a more compact way without  modifying the 
Green functions of the physical fields.
The 1PI Green functions of the theory are obtained differentiating $\g$   
with respect to some of the SM fields. Differentiation of   \equ{nielsen}  
therefore gives  the gauge-dependence of a Green function
in terms of products of other Green functions, 
which also contain the source $\chi$. 

We denote by $\g^{(n)}_{\varphi_1\varphi_2,...}(p_1,p_2,...)$ 
the 1PI Green function
of  $\varphi_1,\varphi_2,...$ at the $n$-loop level.  
$\varphi_i$ can be  any physical or unphysical field of the SM in a general
covariant $R_\xi$ gauge, 
as well as any of the sources $\gamma_{\varphi_i}$, $\chi_j$ 
associated to the BRST variation of $\varphi_i$ and  of the gauge parameter
$\xi_j$. $\g_{\varphi_1\varphi_2...}$ can be expressed as  functional
derivatives of the generating functional, the effective action $\g$, with
respect to the fields and sources $\varphi_1,...,\varphi_m$,
\be
\g_{\varphi_1...\varphi_m}(p_1,...,p_m)=\left.
\frac{\delta^m \g}{\delta \varphi_1(p_1)
\, ... \, \delta\varphi_m(p_m)}\right|_{\varphi_i=0}.
\nonumber\ee
The exchange of two fermionic indices leads to a change
in sign.
We also adopt the short-hand notation
 $\de_\xi$ for the partial derivative with respect to a generic 
gauge parameter $\xi$, whose associated source is generically called $\chi$. 
Some details concerning the action of the 
Slavnov-Taylor operator ${{\cal S}_\Gamma}$ on $\g$, the precise gauge-fixing
term, and the complete source Lagrangian are given in  App.\,A. 
Notice that $\Gamma$ and its Green functions are renormalized objects, 
unless explicitly stated.

\begin{comment}{
In the most general case, the renormalization procedure at order $n$ introduces
several  modifications to \equ{nielsen}.
First, the renormalization at order $n-1$ of the physical 
parameters of the SM  can 
induce additional gauge-dependence at order $n$ if the renormalization 
conditions are not chosen accordingly (we will see a few examples of that
 in the following). Second, the renormalization of
the fields and of the unphysical sector may deform the Nielsen identities. 
}\end{comment}

Before we consider the quantum counterpart of \equ{nielsen},
it is necessary to discuss the parametrization of the theory in some detail.
We  distinguish between three different categories of renormalization 
conditions.
\begin{itemize}
\item [i)] The ones that fix the {\it physical} parameters $p_i$, namely
 the parameters of the 
classical gauge-invariant Lagrangian. They must be fixed using physical 
observables $O_i$ (cross sections, decay rates, resonance parameters etc.):
\be
O_i=f_i(p_j).
\label{rencon}
\ee 
A set of renormalization conditions  commonly used in  
precision calculations is given by
the fine structure constant $\alpha$, the Fermi constant $G_F$ (measured in 
the muon decay), $\alpha_s$ (measured
e.g.\ from the ratio $R$ of hadronic to leptonic decays of the $Z^0$), 
the mixing parameters of the quark sector 
(measured e.g. in hadronic decays),
and the masses of the  $Z^0 $, the Higgs boson $H$,  and  all the fermions. 
In order to keep the renormalization program simple, it is indeed standard procedure 
to adopt mass parameters also for unstable fields. This has the advantage 
of establishing a direct connection between an experimental
quantity  and the two-point Green functions, $\g_{\phi\phi}(q^2)$. 
On the other hand, 
the identification of the masses of unstable particles 
from the resonance parameters is not straightforward
beyond the lowest orders of perturbation theory. 
For example, the masses of unstable particles are often defined
in terms of the zero of the real part of the two-point function, i.e.\
by imposing
\be
{\rm Re}\, \g_{\phi\phi}(M^2_\phi)=0.
\label{conv}
\ee
This definition is not gauge independent beyond one-loop \cite{zmass,psw}
unless $\phi$ is a stable field, 
but it has been used sometimes also in all-orders analyses \cite{kraus}.
A similar problem of gauge-dependence may arise if one tries to define
the  mixing parameters in the quark sector in terms of two-point functions
only, instead of relying on physical amplitudes \cite{ckm}. It follows that,
 if physical amplitudes are not {\em directly} employed like in \equ{rencon},
the consistency 
of the renormalization  conditions has to be proved by means of the Nielsen 
identities, to all orders in perturbation theory, and the connection
between theoretical constructs and experimental quantities has to be
elucidated. 
In our discussion we will fix all the physical parameters using  
unambiguously defined physical observables (cross-sections, decays rates etc.).
All sets of physical observables are equivalent and are  chosen  
according to the experimental precision of the inputs and to the convenience
for the  problem at hand. It is irrelevant for our analysis which set is
actually employed. Having defined the physical parameters in terms of
observables,  we will show that 
the position of the complex pole of the propagator of all physical fields 
of the SM is a gauge independent quantity and can be used to define the mass
parameters, provided the connection between  the field-theory concept and the
experimental quantities is clarified. This is the case, for instance,
for the $Z^0$ mass parameter defined from the complex pole, whose
relation with the resonance shape measured at LEP is well-understood --- 
see the first of Refs.\,\cite{zmass}.
This procedure applies also in the case of mixing between different fields.

\item [ii)] The conditions needed to prevent infrared (IR) divergences.
Due to the presence of massless degrees of freedom, 
it is necessary to impose some
auxiliary conditions that guarantee the correct IR behaviour of the theory.
In particular, it is necessary to impose $\g_{\smalla\smallz}(0)=0$ and similar
conditions in the ghost sector \cite{BFM,kraus}.
  
\item [iii)] Other unphysical renormalization conditions, such as wave 
function  renormalizations, tadpole and gauge parameter renormalization. 
Apart from the case of the tadpole, discussed in Sec.\ref{tadpole}, we do not 
restrict ourselves to a specific choice, but simply require that they do 
{\it not} spoil  the STI and do not  affect the nilpotency of the
Slavnov-Taylor operator. An alternative approach is followed in \cite{kraus}.
\end{itemize}

We recall that
no invariant regularization is known for the SM. The 
implementation of dimensional regularization of 
\efe{dieter}, for instance,  is consistent but breaks the STI. 
They have to be restored  order by order through the introduction of 
non-invariant counterterms --- see e.g.\ \cite{sorella,ghs}.
This is a precondition to any discussion of the renormalization
and it is necessary to recover the unitarity of the theory and the physical
interpretation of the S-matrix amplitudes. 

Unlike the STI, \equ{nielsen} does not have to be preserved in the 
renormalization process, as the extended BRST symmetry is just a technical 
tool for the derivation of the Nielsen identities. Therefore, in the 
following we will consider the possible deformations of \equ{nielsen} 
 induced by quantum effects in complete generality\footnote{In the case of 
Yang-Mills theories, 
a discussion of the renormalization of the Nielsen Identity  
can be found in Ref.[2]; it agrees with the one given below.} 
and write: 
\be
\de_\xi %
\g =  {{\cal S}_\Gamma} \left( \frac{\de}{\de \chi}\,\g    \right) +
\Delta
\label{nielsen2},
\ee  
where the symmetry breaking term $\Delta$ \,is a dimension four operator 
with zero ghost number such that  ${\cal S}_\Gamma \,\Delta=0$.

The investigation of the structure of $\Delta$ in \equ{nielsen2} 
can be performed according to standard cohomological techniques 
\cite{sorella,kraus,BFM}. 
Recalling that ${\cal S}_\Gamma^2=0$ if ${\cal S} \left( \g \right) =0$,
the first step consists in writing $\Delta=X+ {\cal S}_\Gamma Y$ with 
$X\neq {\cal S}_\Gamma \Xi$. As can be intuitively understood, the part of 
$\Delta$ which can be expressed as the 
BRST variation of something else does not contribute to physical quantities.
On the other hand, $X$ does not decouple from the calculation of observables  
and  is usually called the cohomology of the operator ${\cal S}_\Gamma$.
In the SM, $X$   is  composed  of   the dimension four 
gauge-invariant operators with zero ghost number, 
each of them representing a cohomology class\footnote{We recall that 
in the SM, besides the STI, some auxiliary constraints are needed 
to identify the gauge invariant 
operators. For a detailed discussion we refer to \cite{henri,kraus,BFM}.}.
The coefficients of the cohomology classes  of ${\cal S}_\Gamma$
are the physical parameters of the theory.
 Therefore, a contribution to $X$ 
can be absorbed into a  renormalization of some of the physical 
parameters $p_i$ and
we  can write $X=\sum_{i} \beta_i^\xi \,\frac{\de}{\de p_i} \g$.
 For what concerns ${\cal S}_\Gamma Y$, 
it admits different kinds of  contributions 
and is extensively studied in the literature \cite{sorella,dieter2,low-sh}.
The most general expression for (\ref{nielsen2}) turns out to be
\be
\de_\xi %
\g = (1+\rho^\xi) \,
{{\cal S}_\Gamma} \left(  \frac{\de}{\de \chi}\,\g    \right)
+ \sum_{i} \beta_i^\xi \,\frac{\de}{\de p_i} \g +
\sum_\varphi \gamma_\varphi^\xi \,{\cal N}_\varphi \,\g +
\delta_t \int d^4 x \frac{\delta \g}{\delta H(x)}
\label{nielsenren}.
\ee  
In App.\,B we show how this structure is actually implemented and
preserved at all orders.
In \equ{nielsenren}   $p_i$ are the renormalized parameters of the SM,
  $\beta_i^\xi $ describes their explicit gauge dependence 
 (or equivalently that of their corresponding counterterms),
and $\varphi$ is any of the physical or unphysical fields of the SM.
When \equ{nielsenren}
is differentiated to obtain identities between  Green functions,
the operator ${\cal N}_\varphi$ counts the external fields, 
while $\rho^\xi$, $\gamma_\varphi^\xi$ and $\delta_t$ 
parametrize the deformation of the Nielsen identity; they correspond
to a renormalization of unphysical parameters. In particular, 
the third term in \equ{nielsenren} renormalizes the external fields 
(wave function renormalization), the fourth renormalizes the tadpoles, and 
$\rho^\xi$ rescales the gauge parameters. As in the SM 
with restricted  `t~Hooft gauge-fixing there 
are four gauge-fixing parameters $\xi_i$ ($i=Z,W,\gamma,g$) and as many sources
$\chi_i$,  $\rho^\xi$
is in fact  a matrix. In the case of mixing between fields characterized by
the  same quantum numbers, 
$\gamma_\varphi^\xi$ and ${\cal N}_\varphi$ are also matrices. 

\equ{nielsenren} shows the most general structure of the renormalized
Nielsen identity. In many practical cases, however,
the situation is simpler. 
First, our assumption on the renormalization of the physical parameters
in terms of physical quantities implies automatically $\beta^\xi_i=0$. 
This follows directly from \eqs{rencon}, as $O_i$ are gauge-independent
physical objects.
If the physical 
renormalization conditions were mistakenly chosen in a gauge-dependent
way, non-vanishing $\beta^\xi_i$ would arise because spurious gauge-dependence 
would be introduced in \equ{nielsen}.

In pure QCD, where  naive dimensional regularization is consistent and
respects the STI, it is customary to adopt a minimal subtraction ($\ms$) 
as an intermediate renormalization condition\footnote{This is also common in 
some one and two-loop \ew calculations \cite{msbar}.}. Such  $\ms$ subtraction
leads  in this case not only to
$\beta_i^\xi =0$, because the renormalized parameters are guaranteed 
to be gauge-independent \cite{wilzcek}, but also to $\Delta=0$.

One can also wonder whether the NI can be realized
at all orders by an appropriate set of unphysical renormalization
conditions. Although a complete analysis of this problem is beyond the
scope of the present paper, the possibility of  
preserving the form of the NI without modifying {\it ad hoc} the structure
of the ST operator (as in \cite{si-pig}) seems unlikely \cite{kraus-si}. On the other hand, 
the point of view we have followed here has been to allow for 
arbitrary deformations of the NI.

The decomposition of $\Delta$ in \equ{nielsen2} into $X$ and ${\cal S}_\Gamma
Y$ becomes important in the calculation of physical observables.
Since any operator that
 can be expressed as the BRST  variation of something else decouples 
from physical quantities,  ${\cal S}_\Gamma
Y$ is completely irrelevant to their calculation. Hence, no contribution
to the third, fourth and $\rho_\xi$ terms in the rhs of \equ{nielsenren} 
has an effect on physical quantities.
In Sec.\,\ref{ampl} we will consider, in particular, the gauge cancellations 
leading to gauge-independent physical amplitudes. 
\equ{nielsenren} tells us that neither 
the  renormalization of the fields, nor the one of the unphysical parameters, 
can spoil  the gauge independence of the amplitudes.
Only $X=\sum_{i} \beta_i^\xi \,\frac{\de}{\de p_i} \g$ can make them  gauge 
dependent \cite{brs}. In other words, only the renormalization of the 
physical  parameters of the theory affects the gauge-dependence
of the physical observables.

\section{Tadpoles} 
\label{tadpole}
As a  preliminary step in our analysis, we consider in this section the
gauge-parameter dependence of the tadpoles. This is a very simple case and
provides a first introduction to the use of the Nielsen identities.
Because of the close connection between the  mass and the
tadpole renormalizations, the results of this section will be necessary
in all subsequent applications.

The 1PI generator $\g$ is defined as the Legendre transform of the connected 
generating functional ${\bf Z}$. 
The Legendre transform is  well-defined only if   the linear terms in the 
fields (tadpoles)  are  removed  at all orders in perturbation theory 
\cite{itz,becchi}.
This is equivalent to setting  the  renormalization condition
\be
\g_\smallh^{(n)}=0,
\label{notad}
\ee
and  also  corresponds to minimizing the effective 
potential at each order \cite{taylor}\footnote{
Incidentally, it is interesting to see that the tadpole counterterm 
is generated by the BRST variation of a local counterterm:
$\g^{CT} = \delta 
T \, {{\cal S}_{\Gamma_0}} \left( \int \dx \,\gamma_\smallh \right) = \delta T
\int \dx  \, \frac{\del \g_{0} }{\del H(x)}, $
where $\g_{0}$ is the tree level action and $\delta 
T$  the coefficient of this counterterm.  
It then follows  that a renormalization of the
tadpole amplitude induces a shift proportional to $\delta T$ in the mass 
parameters of all the SM fields. 
The previous equation  uncovers also the unphysical nature 
of the renormalization of the tadpole.}. 

We  now consider how the  condition of \equ{notad} affects the  Nielsen Identity.
First, we differentiate  
both sides of \equ{nielsenren}
with respect to $H$. Taking into account \equ{app3} and setting all 
deformation 
parameters to zero, but before employing \equ{notad}, we obtain 
\be
-\de_\xi \g_{\smallh} (0)= \g_{\chi \gamma_{\smallh} \smallh}(0)\, \g_{\smallh}(0) +
\g_{\chi\gamma_{\smallh}}(0) \,\g_{\smallh\smallh}(0).
\label{tad1}
\ee
All the external momenta are zero and  we will drop them in the following of
this section. As $\chi$ is the source associated to a gauge parameter, 
it is a Grassman variable which does not depend on the 
space-time and does not carry any momentum. 
In deriving \equ{tad1}, we have used the fact that 
the  $\chi$'s and the $\gamma$'s have  ghost number equal to +1 
and $-1$, respectively, and that non-vanishing Green functions must conserve 
the ghost charge.     We have also used CP  
conservation to avoid, for instance, the appearance of
$H$-$G^0$ mixing in higher orders. This assumption will be relaxed later.

The renormalization of $\g_{\chi\gamma_{\smallh}}$, which is logarithmically divergent,
has to be fixed  explicitly. It  follows from \eqs{tad1} and (\ref{notad})  
 that for the Nielsen identity  not to be  deformed we must impose  
\be
\g_{\chi\gamma_{\smallh}}^{(n)}=0
\label{constraint}
\ee
at any order $n$ of perturbation theory.
If we allow the renormalization of $\g_{\chi\gamma_{\smallh}}$ to  
deform  Eq.\,(\ref{nielsenren}) according to \equ{nielsenren}, however,
we have 
\be
(1+\rho^\xi)\,\g_{\chi\gamma_{\smallh}} + \delta_t \, \g_{\smallh\smallh}(0)=0.
\ee
In the following we will consistently impose \equ{constraint}. 

In the presence of CP violation, another tadpole amplitude emerges in the SM,
connected to the vacuum expectation value of the CP-odd neutral would-be 
Goldstone boson,
$G_0$. As the CP violation in the SM is confined to the fermionic sector, 
this will happen only at higher orders. In extended models,
any neutral scalar 
field with zero ghost charge could develop a vacuum expectation value
through radiative corrections. In all cases the linear terms in the fields must be
removed. However, given \equ{notad}, the STI imply the vanishing of tadpoles
of the unphysical fields. 
Upon differentiation with respect to the neutral ghost field $c^{\smallz}$, 
\equ{app3} yields 
\begin{eqnarray}\label{tad3}
\left.\frac{\del \,{{\cal S}_\Gamma} 
\left(  \g \right)}{\del c^{\smallz}(0)}\right|_{\varphi=0}  = 
\g_{c^\smallz \gamma_0} \,\g_{\smallG_0} + \g_{c^\smallz \,\gamma_{\smallh}}
 \g_\smallh=0.
\end{eqnarray}  
To derive the previous equation, we have used \eqs{app3} and (\ref{app4})
and the fact that  one-point functions are not vanishing only for 
neutral scalars with zero ghost number. As can be seen from 
\equ{app4}, $\g_{c^\smallz \gamma_0}^{(0)}$ differs
from zero already at the tree level, in which case it is proportional to $v$,
the  Higgs v.e.v..  From \equ{tad3} it then follows that the  vanishing of the 
CP-even tadpole $\g_\smallh^{(n)}$ implies the vanishing of the 
CP-odd tadpole $\g_{\smallG_0}^{(n)}$ at any order.
Moreover, 
in the presence  of CP violation a term $\delta_t^{CP} \int d^4x \,\delta\g/
\delta G_0(x)$  should be added to \equ{nielsenren}.
Using the STI for the two-point functions and  the analogous 
of \equ{tad1}, and requiring $\delta_t^{CP}=0$ one  then finds that
\equ{constraint} is also valid, together with $\g_{\chi\gamma_0}^{(n)}=0$.

In the case of a model with two Higgs-doublets \cite{2HDM}, 
\equ{tad3} takes the form
\begin{eqnarray}\label{tad4}
\g_{c^\smallz \gamma_0} \,\g_{G_0} + \g_{c^\smallz \,\gamma_{\smallh}}
 \g_\smallh + \g_{c^\smallz \,\gamma_{h}}
 \g_h + \g_{c^\smallz \,\gamma_{A}}
 \g_A=0.
\end{eqnarray}  
where $H,h$ and $A$ are the physical neutral Higgs fields.
It is sufficient to     
require only the vanishing of the tadpoles of the physical fields $H,h,A$. 
It then  follows that the tadpole of  the unphysical 
Goldstone boson is zero
(identifying a {\it flat direction} in the Higgs potential \cite{apostolos})
at any order in perturbation theory.

\begin{figure}[t]
\centerline{
\psfig{figure=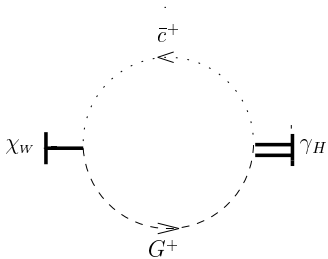,height=5.in,rheight=0.8in}  }
\caption{\sf One-loop diagram contributing to $\g_{\chi\gamma_\smallh}$.}
\label{figtad}
\end{figure}

Before closing this section, 
it is instructive to  check explicitly \equ{tad1} at the one-loop order.
At this order none of the pathologies of (naive) dimensional regularization
is apparent and we have an elementary  example of a 
calculation with the $\chi$ sources.
To this end we expand \equ{tad1}  at $O(g)$ and consider dimensionally 
regularized Green functions before implementing the renormalization 
conditions.  
As a consequence of the Feynman rules given in  App.\,A,
the tree level Green functions
$\g_{\chi \gamma_{\smallh} \smallh}^{(0)}$ and 
$\g_{\chi\gamma_{\smallh}}^{(0)}$
vanish.  We also have $\g_{\smallh}^{(0)}=0 $ by construction, while 
$\g_{\smallh\smallh}^{(0)}(0)=-\mh^2$. We therefore find
\be
\de_\xi \g_{\smallh}^{(1)}= \mh^2
\,\g_{\chi\gamma_{\smallh}}^{(1)}, 
\label{tad2}
\ee
where the last term  is logarithmically divergent. It is straightforward to 
compute $\g_{\chi\gamma_{\smallh}}^{(1)}$ using the Lagrangian given in 
App.\,A. Only diagrams of the kind displayed in Fig.\ref{figtad} contribute
and we recover the gauge dependence of  $\g_{\smallh}^{(1)}$
given in Eqs.(11,12) of \cite{ds92}.

\section{W boson}
As a first application of the technique to the case of the definition of mass,
we consider the case of the  charged 
$W$ boson, which  is  particularly simple 
because it does not involve any mixing between different fields.
We split the  inverse $W$ propagator    into its 
transverse and longitudinal parts
\be
\g^{\mu\nu}_{\smallw\smallw}(q)= 
\left(g^{\mu\nu} -\frac{q^\mu q^\nu}{q^2}\right) 
\g^{{\scriptscriptstyle T}}_{\smallw\smallw}(q^2)
+ \frac{q^\mu q^\nu}{q^2} \g^{{\scriptscriptstyle L}}_{\smallw\smallw}(q^2).
\ee
Our first aim is to obtain a Nielsen identity for the transverse part of the
two-point function.
The longitudinal part will be considered in Sec.\,\ref{scalar}.
As a first step, we differentiate
both sides of \equ{nielsen} with respect to $W_\mu^+$ and 
$W_\nu^-$, take into account \equ{app3}, and   set to zero the Green
functions which do not conserve the ghost charge. We obtain
\bea
\partial_\xi  \g_{\smallw\smallw}^{\smallT}(q)&=&
-\dms{\sum_\varphi} \left[
\g_{\chi\gamma_{\varphi} \smallw\smallw}^\smallT (q)\,\g_\varphi +
\g_{\chi\gamma_{\varphi}}\, \g_{\varphi\smallw\smallw}^\smallT(q)\right.\non\\
&&\ \ \ \ \ +\left. t^{\mu\nu} \left(
 \g_{\chi\gamma_{\varphi}\smallw_\mu}(q) \,\g_{\varphi\smallw_\nu}(q)+
 \g_{\chi\smallw_\nu\gamma_{\varphi}}(q) \,\g_{\smallw_\mu\varphi}(q)\right)
\right]
\label{w1}
\eea
where $t^{\mu\nu}=g^{\mu\nu}-{q^\mu q^\nu}/{q^2}$ is the 
transverse projector and the superscript $T$ indicates the transverse part of a
Green function. 
From the discussion of the previous section we know that there is no
non-vanishing one-point function and that 
$\g_{\chi\gamma_{\varphi}}$ for $\varphi=H,G_0$, which describe the
gauge-dependence of the tadpoles, must also vanish --- see \equ{constraint} ---
if we impose  $\delta_t=\delta_t^{CP}=0$.
 The second line of \equ{w1}, on the other hand, 
is not zero only for $\varphi=W^\pm_\lambda$, so that we obtain,
 at any order in perturbation theory ($s=q^2$),
\bea
\partial_\xi  \g_{\smallw\smallw}^{\smallT}(s)&=
&%
- 2\,\g_{\chi\gamma_{\smallw}\smallw}^\smallT(s) \,
\g_{\smallw\smallw}^\smallT(s)
\label{w1b},
\eea
with $\g_{\chi\gamma_{\smallw^+}\smallw^-}^\smallT=
\g_{\chi\gamma_{\smallw^-}\smallw^+}^\smallT$.
We now include the possible deformations present in \equ{nielsenren}:
using $\beta_i^\xi=0$, \equ{w1b} becomes 
\be
 \partial_\xi  \g_{\smallw\smallw}^{\smallT}(s)=
2\left[- \left(1+\rho^\xi\right)\g_{\chi\dms{\gamma_{\smallw}}\smallw}^\smallT(s) + 
\gamma_\smallw^\xi\right]  \,
\g_{\smallw\smallw}^\smallT(s).
\label{w2}
\ee
For what concerns the mass parameter definition, 
the significance of \equ{w2} is that a
gauge invariant  and self-consistent normalization condition on 
$\g_{\smallw\smallw}^{\smallT}(s)$ can {\it only} be given at the location 
of the pole of the propagator. Defining the latter by
\be 
\g_{\smallw\smallw}^{\smallT}(\bar{s}_\smallw)=0
\label{w3},
\ee
we see that \equ{w2} leads to $\de_\xi
\g_{\smallw\smallw}^{\smallT}(s)|_{s=\bar{s}_\smallw}= \de_\xi
(\g_{\smallw\smallw}^{\smallT}(\bar{s}_\smallw))$, which in turn 
implies that {\it the location $\bar{s}_\smallw$ of the complex
pole of the propagator is gauge-independent at any order in perturbation
theory}. This is a remarkably non-trivial result of 
 perturbation theory, as it concerns  the parameters that
describe the  non-perturbative phenomenon of resonance. 
It relies exclusively on $\beta_i^\xi=0$, which follows from
our use of observables to fix all the physical parameters.
The mass parameter $m_\smallw$ and the width parameter $\Gamma_\smallw$ 
defined by ${\bar s}_\smallw=m_\smallw^2 -i \,m_\smallw \Gamma_\smallw$
are gauge independent quantities and, as a consequence of the discussion 
at point (i) in Sec.\,2, $m_\smallw$ 
can be adopted as renormalized $W$ mass. Clearly, the precise connection 
between this parameter and related experimental quantities must be clarified
in order to adopt \equ{w3} as a renormalization condition that directly
fixes the $W$ mass parameter.

\begin{figure}[t]
\centerline{
\psfig{figure=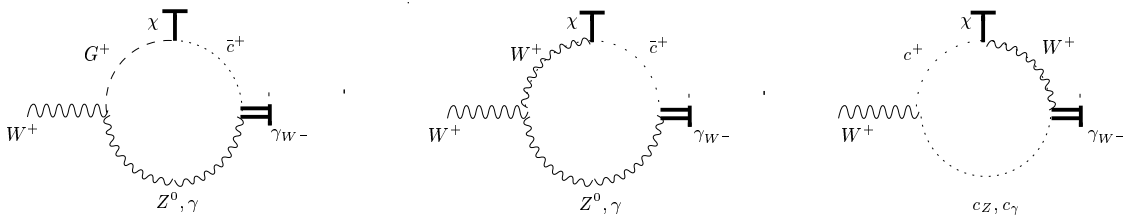,height=7.1in,rheight=0.9in}  }
\caption{\sf One-loop diagrams contributing to $\g_{\dms{\chi_\smallw
\gamma_\smallw} \smallw}$.}
\label{fig2}
\end{figure}

Beyond one-loop order the definition of the mass parameter of an unstable
particle in terms of its two-point function is not trivially gauge-independent
as in \equ{w3} \cite{zmass,psw}.
Consider for instance the case in which the mass of the $W$ boson is defined 
by a renormalization condition of the kind in \equ{conv}, namely 
\be
{\rm Re}\,\g_{\smallw\smallw}^{\smallT}(\mw^2)=0
\label{w4};
\ee
the  $W$ mass counterterm is then  ${\rm Re}\,
\g_{\smallw\smallw}(\mw^2)$. This is the conventional approach 
to one-loop mass renormalization \cite{si80,aoki,hollik}. 
Taking the real part of \equ{w2} at $s=\mw^2$, expanding it  at
two-loop, and dropping $\rho^\xi$ and $\gamma^\xi_\varphi$
as they would not affect our conclusions being real,  we obtain
\bea
\de_\xi {\rm Re}\,\g_{\smallw\smallw}^{\smallT(2)}
&=& -2\,{\rm Re}\,\g_{\chi\gamma_{\smallw}\smallw}^{\smallT(1)}
{\rm Re}\,\g_{\smallw\smallw}^{\smallT (1)}
+2\, {\rm Im}\,
\g_{\smallw\smallw}^{\smallT (1)}\ {\rm Im}\,\g_{\chi\gamma_{\smallw}
\smallw}^{\smallT(1)}\nonumber
\eea
where all terms are evaluated at $q^2=\mw^2$.
Using the normalization condition \equ{w4}, we see that the last term 
is left over, so that
\equ{w2} is not satisfied. As a consequence, the mass parameter defined by 
\equ{w4} is gauge-parameter dependent beyond one-loop \cite{psw}. 
As the  imaginary part in the last term of the previous equation
originates from gauge-dependent thresholds, there  exists a class of 
gauges where  ${\rm Im}\,\g_{\chi\gamma_{\smallw}\smallw}^{\smallT(1)}
(\mw^2)$ vanishes (Cf. Fig.\,\ref{fig2}) and  for which the gauge 
parameter dependence of $\mw$ 
is only apparent at the three loop level \cite{psw}. 
The actual difference between the two mass definitions, 
$\Delta M^2=
{\rm Re} [\g_{\smallw\smallw}(\bar{s}_\smallw)-\g_{\smallw\smallw}(\mw^2)]$,
can be evaluated  expanding $\g_{\smallw\smallw}$ in powers of
$|\bar{s}_\smallw-\mw^2|\approx\Gamma_\smallw \mw=O(g^2)$ up to $O(g^4)$.
The result is $\Delta M^2\approx\mw \Gamma_\smallw {\rm Im} \,
\g_{\smallw\smallw}^{(1)'}(\mw^2)$, which %
is clearly gauge parameter dependent. 
The renormalization condition (\ref{w4}) is an %
example of definition of a physical parameter in a gauge-dependent way:
beyond one-loop it induces $\beta^\xi_{\mw}\neq 0$.

A comment on the factor $\gamma^\xi_{\smallw} $ is now in order.
As explained in the introduction, this term originates from the 
potential deformation of the Nielsen identity by 
the renormalization procedure. 
For instance, there is considerable freedom in the choice of 
both the wave function renormalization of the $W$ field and 
the renormalization of 
$\g_{\chi\gamma_{\smallw}\smallw}^\smallT(s)$.
In case they do not respect the Nielsen identity,
$\gamma^\xi_{\smallw} $ compensates for its breaking. 
Let us consider, for ex.,   the following two procedures at one-loop. 
A first possibility is   
to adopt a minimal subtraction ($\ms$ scheme) for both
the wave function renormalization of the $W$ and 
$\g^{\smallT,(1)}_{\chi\gamma_{\smallw}\smallw}(s)$.
It should be clear that in this case $\gamma^{\xi,(1)}_{\smallw}=0$.
A second possibility consists in using the on-shell scheme for the $W$ field
rescaling. If we now insist in using a minimal subtraction for 
$\g_{\chi\gamma_{\smallw}\smallw}^\smallT(s)$, \equ{w1b} is 
not satisfied by the finite parts of the counterterms,
leading to a factor $\gamma^{\xi,(1)}_{\smallw}= 
\g^{\smallT,(1)}_{\chi\gamma_{\smallw}\smallw}(\mw^2)|_{\msbar}=\frac1{2}
\g^{\smallT',(1)}_{\smallw\smallw}(\mw^2)|_{\msbar}$, where the subscript 
$\ms$ means that  only the finite part of this Green function is considered.
Similar considerations  apply to $\rho^\xi$, which appears first at the
two-loop level and is related to the renormalization of the
gauge-fixing parameters.

Like in the case of the tadpole, let us see explicitly what happens at the 
 one loop level for regularized Green functions. Using \equ{tad2}
and noting that the Green functions involving $\chi$ vanish at the tree level,
\equ{w1}  reduces to 
\be
\partial_\xi  \left[\g_{\smallw\smallw}^{\smallT,(1)}(s) +T^{(1)}_\smallw
\right]= 2\,\g_{\chi\gamma_{\smallw}\smallw}^{\smallT,(1)}(s) \,
(s-\mw^2),
\label{w1loop}
\ee
where $T^{(1)}_\smallw$ is the contribution of the 
one-loop tadpole. The zero of the $W$ inverse propagator is 
gauge-independent at  $s=\mw^2$. Notice that $\g_{\chi\gamma_{\smallw}
\smallw}^{\smallT,(1)}(s)$ describes the gauge-dependence of the residue of the
physical pole, i.e.\ of the on-shell wave function renormalization factor.
An explicit calculation of the 
diagrams in Fig.\ref{fig2} which contribute to
$\g_{\dms{\chi_\smallw\gamma_{\smallw}}\smallw}^{\smallT,(1)}$ 
leads to the same
$\xi_\smallw$-dependence of $A_{\smallw\smallw}^{(1)}$ reported in \cite{ds92};
 the same happens for the $\xi_{\smallz,\gamma}$-dependence.

We have seen that if the renormalization condition is not properly chosen,
the mass parameter is gauge-dependent.
A possible source of confusion, however, is the interplay of
mass and tadpole renormalization. To make this point clear, it is sufficient
to keep the discussion at the one-loop level. 
From \equ{w1loop} we know that the $W$ mass counterterm 
$\delta\mw^2= {\rm Re} \g_{\smallw\smallw}^{\smallT(1)}(\mw^2) +T_\smallw^{(1)}$
is gauge-independent. The tadpole renormalization according to 
Sec.\,\ref{tadpole}, however, eliminates $T_\smallw^{(1)}$ from the previous 
expression and makes $\delta\mw^2$ gauge-dependent.
Nevertheless,  we still have $\beta^\xi_{\mw}= 0$. 
This is a consequence of the unphysical character of the tadpole 
renormalization. 
What is  essential  here is that the renormalization condition which
fixes the physical parameter $\mw$ be gauge-independent, as is the case for 
\equ{w3} and not for \equ{w4}. 
This and only this  guarantees $\beta^\xi_{\mw}=0$.

Two simple practical applications follow from \equ{w2}, and we report them
as illustrations of the technique. First, we can
consider the dependence of the $W$ self-energy on the QCD gauge-parameter
$\xi_g$. It is easy to show that the deformation parameters cannot affect it
in this case, and that it
is controlled by $\g^\smallT_{\chi_g \gamma_\smallw \smallw}$ only. 
However, the ghost charge associated to the QCD gauge group  and the one
associated to the SU(2) group must be conserved independently of each other.
Therefore,   $\g^\smallT_{\chi_g \gamma_\smallw \smallw}=0$ at any order,
which implies that the $W$ two-point function does not
depend on the gluon gauge parameter, %
as verified in actual calculations at two and three loops \cite{wqcd}.
The second  application concerns the contributions to the 
$W$ self-energy which are leading in an expansion in the heavy top quark mass.
 At the one-loop level, they are trivially gauge-independent, like all
the fermionic contributions. At higher orders, 
one can use the fact that $\g^\smallT_{\chi\gamma_\smallw \smallw}(s)$
is only logarithmically divergent to show that the gauge dependence
of the heavy top expansion of $\g_{\smallw\smallw}$ starts at the
next-to-leading order. Again, this is not surprising, because the leading
contributions in $\mt$ can be obtained in the framework of a
Yukawa Lagrangian where the heavy fermions only couple to the Higgs boson and
to the longitudinal components of the gauge bosons.  This Lagrangian, which
corresponds to the {\it gaugeless limit} of the SM \cite{bar},
does not require  gauge-fixing.

\subsubsection*{Infrared  finiteness of the $W$ mass}
The complex pole definition of mass based on \equ{w3} avoids also IR 
problems at higher orders in perturbation theory.
It has been shown in 
Ref.\,\cite{psw} that the use of the normalization condition of \equ{w4}
leads to severe IR divergences in a class of higher order graphs 
containing the photon when the external momentum approaches 
the mass-shell of the $W$. As a consequence, 
in the resonance region, $|s-\mw^2| \lsim \mw \Gamma_\smallw$, the 
perturbative series fails to converge, while it was found that
the pole mass definition avoids all these pathologies.  
The origin of the problem is  similar to the one of the
gauge-dependence of the mass parameter defined by \equ{w4} 
and is related  to the need to take into account the
imaginary part of $\g_{\smallw\smallw}^\smallT$ in the renormalization procedure.

More generally, the problem is common to all particles coupled to massless 
quanta, independently of whether they are stable or not,
and concerns the perturbative description of the resonance region. 
For instance, in  pure QCD it is well-known \cite{tarr} that at two-loop order
the two-point function
of a massive quark is IR divergent at $q^2=m_q^2$ unless the quark 
mass is renormalized on the pole. In \efe{kronfeld}
it was shown that this property persists at all orders in QCD, namely that
the perturbative pole mass of the quark in QCD is infrared safe (or finite).  
In the following we would like to approach the case of the $W$ boson from a 
slightly different point of view, along the lines of \cite{kronfeld}, 
generalizing some of the results of \efe{psw}.
We will show that the complex pole mass of the $W$
is IR safe at any order in perturbation theory, namely that the renormalization
condition of \equ{w3} does not lead to IR divergences in the resonance region
of the $W$ boson, nor to pathologies in the perturbative expansion.
 In that respect, the presence of the width does not alter the discussion 
in  a relevant way.

\begin{figure}[t]
\centerline{
\psfig{figure=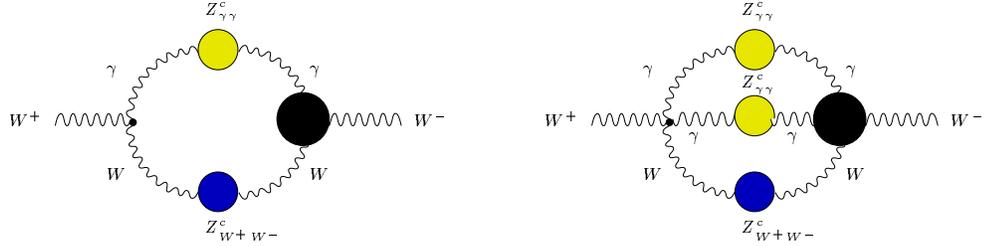,height=9.in,rheight=1in}  }
\caption{\sf Schwinger-Dyson equation for the $W$ two-point function. 
The  blobs on the internal lines represent  connected propagators (chains
of bubbles), while the  blob on the vertex represents  a one-particle
irreducible Green function.}
\label{figir}
\end{figure}

A convenient tool to  analyze the IR properties of the 
$W$ self-energy from a perturbative point of view 
are the renormalized Schwinger-Dyson equations (see e.g.\ \cite{itz}).
 These equations 
provide a  simple iterative way to define the higher order graphs in 
terms of  sub-diagrams.  In the case of the $W$ boson there are only two 
topologies containing the photon which should be considered, as they contain 
thresholds at $s=\mw^2$ and  can lead  at higher orders to IR  problems. 
 Their Schwinger-Dyson equations are
 graphically depicted in  Fig.\,\ref{figir}. 
Diagrams with gauge-dependent threshold (like  those with a charged Goldstone 
boson in place of the $W$) and with thresholds far away from the resonance 
region (like those with a $Z^0$ boson instead of the photon) can be discarded
because their expansion around $s=\mw^2$ does not contain non-analytic terms.

We will treat explicitly 
only the case of the  topology on the left side of Fig.\,\ref{figir}, 
as the other diagram  can be discussed along the same lines. 
In this case the Schwinger-Dyson equation has the form
\begin{eqnarray}
\label{sd_1}
\g^{(\gamma)}_{\smallw^+_\mu \smallw^-_\nu}(p) \sim 
\int d^n k  \ 
\g^{(0)}_{\smallw^+_{\mu} A_\rho \smallw^-_{\sigma}}(k, p+k) 
\, Z^c_{\smalla_\sigma \smalla_\beta}(k)
\, Z^c_{\smallw^+_\sigma \smallw^-_\alpha}(k+p) 
\, \g_{\smallw^+_{\alpha} A_\beta \smallw^-_{\nu}}(k, p) ,
\end{eqnarray}
where $\g^{(\gamma)}_{\smallw^+_\mu \smallw^-_\nu}(p)$ is the 
contribution to the self-energy  due to the exchange of a single photon, 
$\g_{\smallw^+_{\mu} A_\rho \smallw^-_{\sigma}}(k, p+k)$ is the 1PI
 vertex, the superscript $(0)$ indicates that the vertex is considered at the
tree level,  and finally 
$Z^c_{\smalla_\sigma \smalla_\beta}(k)$ and 
$Z^c_{\smallw^+_\sigma \smallw^-_\alpha}(k+p)$ are  the  connected 
propagator for the photon and for the $W$ boson, respectively. 
To study the IR behavior of \equ{sd_1} near the mass-shell, 
we now consider the transverse part of the self-energy 
$\g^{(\gamma)}_{\smallw^+_\mu \smallw^-_\nu}(p)$ and  approach the limit 
$p^2 \rightarrow \bar{s}_\smallw$. We expand the 
propagator into the Dyson series of self-energies and 
tree propagators. Concerning the photon line, we 
 recall that a convenient  choice of 
the normalization conditions for the neutral gauge boson sector, i.e.\ 
$\g^\smallT_{\smallz\smalla}(0)=0$, makes 
$\g_{\smalla\smalla}^\smallT(0)$ vanish at all orders (Cf.\ next section). 
Therefore, the photon
propagator  $Z^c_{\smalla_\sigma \smalla_\tau}(k)$ 
is always proportional to $1/k^2$ in the limit $k \rightarrow 0$ and has
IR dimension -2. 

For what concerns the $W$ propagator, the IR divergent contributions are 
related only to the transverse component of 
$Z^{c}_{\smallw^+ \smallw^-}(k+p)$ because the propagator of the 
longitudinal components of the $W$ boson has a gauge dependent pole. 
In the  on-shell limit for the momentum $p$ 
and for $k \rightarrow 0$, the tree level $W$ propagators present 
in the Dyson series  for $Z^{c,\smallT}_{\smallw^+ \smallw^-}(k+p)$  
are linearly divergent. Therefore, expanding 
$Z^{c,\smallT}_{\smallw^+ \smallw^-}(k+p)$ around $k=0, p^2= \bar{s}_\smallw$ 
we  have 
\begin{equation}
\left. 
Z^{c,\smallT}_{\smallw^+ \smallw^-}(k+p) 
\right|_{p^2=\bar{s}_\smallw} 
\stackrel{k \rightarrow 0}{\sim} 
\,\sum_n \left( \frac{1}{2 p\cdot k} \right)^{n+1} 
\left[ \g^{\smallT}_{\smallw \smallw}(\bar{s}_\smallw) \right]^n.  
\label{wline}
\end{equation}
Here we consider only the most dangerous terms, which vanish if and only if
$\g^{\smallT}_{\smallw \smallw}(\bar{s}_\smallw) =0$.  
Under this condition, $\left. 
Z^{c,\smallT}_{\smallw^+ \smallw^-}(k+p) 
\right|_{p^2=\bar{s}_\smallw}$ is at most linearly
divergent in the IR limit. If, on the other hand, \equ{w3} is not satisfied, 
severe IR divergences appear in each order. The situation is not much improved
if we move off the pole position in the resonance region. Indeed, in this case
the $W$ width acts as an  IR regulator in the denominator of 
\equ{wline}, but leads to a series where the  denominator
$1/(s-\bar{s}_\smallw)\approx O(1/g^2)$ spoils the convergence of the 
perturbative expansion  in the resonance region \cite{psw}.

The last information we need concerns the behavior
 of the vertex $ \g^{\smallT}_{\smallw^+  \smalla_\beta \smallw^-}(k, p)$ 
($T$ refers to the transverse components of the $W$ bosons) 
around $k=0, p^2= \mw^2$. By analyticity and dimensional analysis,
the vertex functions can be at most  logarithmically divergent 
 in the limit $k \rightarrow 0$ 
(this can also be verified exploiting the STI together with a proper
use of  the renormalization conditions).
Having  IR dimension -3, it follows by power counting
that  \equ{sd_1}  does not lead to  IR divergences when the 
integral in the internal momentum $k$ is performed around $k=0$. 

In summary, we have seen that the pole mass
of the $W$ boson, defined by \equ{w3}, is IR safe to  all orders in 
perturbation theory and that only if this definition is adopted
 a perturbative description of the resonance region is possible.

\label{W}
%%%%%%%%%%%%%%%%%%%%%%%%%%%%%%%%%
\section{The $Z-\gamma$ system}
\label{gammaZ}
The main difference between the case of the $W$ boson and the one of
the neutral vector bosons is the presence of mixing. We now directly 
use \equ{nielsenren} with  $\beta^\xi_i=0$ and set
$\rho^\xi=0$ for ease of notation (doing otherwise 
would not modify our results).
Following the same steps as in the derivation of \equ{w2}, and keeping
in mind that the abelian vector field does not need a BRST source, we find
for $i,j=A,Z$ 
\bea
 \partial_\xi  \g_{ij}^{\smallT}(s)&=&-  
\left(\g_{\chi\gamma_{3}i}^\smallT(s) -\gamma^\xi_{i3}\right)
\left[ c_\smallw \g_{\smallz j}^\smallT(s) - s_\smallw 
\g_{\smalla j}^\smallT(s) \right] \non\\
&&- \Big[ c_\smallw \g_{i\smallz}^\smallT(s) - s_\smallw 
\g_{i\smalla}^\smallT(s) \Big]
\left(\g_{\chi\gamma_{3}j}^\smallT(s) -\gamma^\xi_{3j}\right)
\label{z1},
\eea
where $\gamma^\xi_{i3}=\gamma^\xi_{3i} $ is the deformation 
induced by the possible mismatch between 
the wave function renormalization matrix $Z_{ij}$ and the
renormalization of $\g_{\chi\gamma_{3}j}$.
We recall that $\g_{ i k}^\smallT(s)$ is a symmetric matrix.
We now consider the quantity
\be
{\cal D}_{\smalla\smallz}^\smallT(s)= \det \left(
\begin{array}{ll}
\g_{\smalla\smalla}^{\smallT}(s) & \g_{\smalla\smallz}^{\smallT}(s) \\
\g_{\smallz\smalla}^{\smallT}(s) & \g_{\smallz\smallz}^{\smallT}(s) 
\end{array}\right),
\ee
which appears in the denominator of the propagators
of the photon-$Z^0$ system (see for ex. \cite{hollik}). 
If we are interested in the analytic structure of
neutral current amplitudes in the typical configuration of a high-energy
collider, where external fermion masses can be neglected, 
${\cal D}_{\smalla\smallz}^\smallT(s)$ is what we need to investigate.
It is straightforward to derive 
\be
\de_\xi {\cal D}_{\smalla\smallz}^\smallT(s) =
-2 \left( c_\smallw \g_{\chi\gamma_{3}\smallz}^\smallT(s)-
c_\smallw \gamma^\xi_{3\smallz}-
s_\smallw\g_{\chi\gamma_{3}\smalla}^\smallT(s)-
s_\smallw\gamma^\xi_{3\smalla}\right)
{\cal D}_{\smalla\smallz}^\smallT(s).\label{det}
\ee
This tells us that the  zeros of ${\cal D}_{\smalla\smallz}^\smallT$
identify gauge-independent quantities. On the other hand, we know from the STI
that ${\cal D}_{\smalla\smallz}^\smallL(0)=0$ (see for ex. \cite{aoki}; 
Ref.\,\cite{BFM} considers also  the case of CP violation) which in turn
 implies by analyticity 
${\cal D}_{\smalla\smallz}^\smallT(0)=0$. This result 
ensures the existence of a massless state, the photon.
${\cal D}_{\smalla\smallz}^\smallT$ has, 
however, another zero, corresponding to
the $Z^0$ complex pole, at $q^2=\bar{s}_\smallz$.
As in the case of the $W$ boson, this result implies that the position of the
complex pole is a gauge independent quantity and that the only 
self-consistent  normalization condition for the $Z^0$ mass
is the one given in analogy to \equ{w3}.
With the exception of the IR problems, all the discussion on the $W$ mass
applies directly to the case of the $Z^0$ boson mass \cite{zmass}.
A Ward Identity similar to the Nielsen identity of \equ{w2} has been applied in
\cite{hveltman} to the case of the $Z^0$ resonance, to the same avail.

Another interesting application of   \equ{z1} concerns the photon 
correlator at $q^2=0$. As is well known \cite{aoki}, 
using the renormalization condition  $\g^\smallT_{\smalla\smallz}(0)=0$ 
the result 
${\cal D}^\smallT_{\smalla\smallz}(0)$ that we have used above implies
$\g_{\smalla\smalla}^\smallT(0)=0$.  
%In this case the only object that enters 
%the standard  electric charge renormalization is 
In this case it is straightforward to verify from \equ{z1} that
the derivative wrt $ q^2$ of the photon two-point function
calculated at $q^2=0$ %that this object 
is gauge-independent at all orders. 
Imposing the condition $\g^\smallT_{\smalla\smallz}(0)=0$ 
in the expression of 
$\de_\xi \g^\smallT_{\smalla\smallz}(0)$, we obtain the constraint
$\g^\smallT_{\chi\gamma_3\smalla}(0) - \gamma^\xi_{3\smalla}=0$. 
We can now differentiate $\de_\xi \g_{\smalla\smalla}^\smallT$ wrt $s$ 
and evaluate it at  $s=0$. Using the various constraints we have obtained, 
we immediately derive 
\be
\de_\xi\, \left.  \frac{\de}{\de{s}}
\g_{\smalla\smalla}^\smallT(s) \right|_{s=0} = 0.
\label{photon}
\ee
Notice that no particular renormalization condition on  the derivative
$\frac{\de}{\de{s}}
\g_{\smalla\smalla}^\smallT |_{s=0}$ has been imposed, so one should think,
 for instance, of a minimal subtraction.
This interesting and  non-trivial result shows that under the condition
$\g^\smallT_{\smalla\smallz}(0)=0$ and at $s=0$ 
there exists in the full SM something  analogous to what happens in QED,
 where the vacuum polarization of the photon is
gauge-independent for any $s$ (see for ex. \efe{bls}).
An alternative derivation of \equ{photon} can be obtained starting from the 
physical photon-electron amplitude at $s=0$,  proceeding along the lines
of the discussion of Sec.\,\ref{ampl}, and taking the gauge-independence of the
on-shell amplitude for granted. 

\section{The scalar sector}
\label{scalar} 
In the previous section we have studied a first example of mixing.
Indeed, mixing occurs in several  other cases in the SM and in most of 
its extensions; all can
be treated in a way very similar to the $\{\gamma, Z\}$ case discussed above.
In this section, we first consider the matrix 
$\g^\phi(s)$ of the two point functions relative to the scalar fields
$\phi=\{\phi_1, \phi_2, ...,\phi_n\}$
in the general case of mixing  and show that the gauge dependence 
of its determinant follows an equation analogous to \equ{det}, if the rank of
$\g^\phi(s)$ is equal  to its  dimension $n$.
As CP violation is present in the SM, 
we then consider the system  formed by $\{A_\smallL,Z_\smallL,G_0,H\}$, 
where  the subscript $L$ denotes the longitudinal component of 
the vector boson fields.
This system is highly constrained by the STI and we 
show that  in this case the complex pole of
the only physical field, the Higgs boson, is gauge-invariant.
In an analogous way one can consider the $\{W_\smallL^\pm,G^\pm\}$ system, 
which however 
has no physical degree of freedom and is completely constrained by the STI.

The general form of the Nielsen identity in the case of a system $\phi$ of
fields characterized by the same conserved quantum numbers can be obtained
in analogy to  \equ{z1} and reads
\be
\de_\xi \g^\phi(s)= \Lambda(s)\, \g^\phi(s) + \g^\phi(s) \,\Lambda'(s),
\label{Ngen}
\ee
where we do not need to specify the matrices $\Lambda$ and $\Lambda'$
any further. Using $\ln\,\det\g^\phi=
{\rm tr}\,\ln \g^\phi$ and exploiting the properties of the trace,
 one finds for ${\cal D}_\phi\equiv \det \g^\phi$
\be
\de_\xi {\cal D}_\phi(s)= {\rm tr} \left[ \Lambda(s)+\Lambda'(s)\right]
{\cal D}_\phi(s),\label{detgen}
\ee
which generalizes \equ{det} in the case the rank of $\g^\phi(s)$
at arbitrary $s$ is equal to its  dimensionality.
In the case of $n$ scalar fields this ensures the gauge-independence of 
$n$ complex poles. Notice that
the physical information contained in the matrix $\g^\phi$ 
is not restricted to the physical poles. Indeed, the higher order
definition of the mixing parameters is  affected by the off-diagonal 
elements of $\g^\phi$. 
In general, it does not seem possible to form 
gauge-independent quantities on the basis of two-point functions only,
i.e.\ of $\g^\phi$,
and to employ them  to renormalize the mixing parameters \cite{ckm}. 
On the other hand, the mixing parameters can be safely defined in terms 
of physical observables such as mesonic decay rates.

Neutral current processes are mediated by photons and  $Z^0$, as
well as by scalar fields, like $G_0$ and the physical Higgs. 
As it is well-known, the propagator matrix is obtained by inversion of 
the two-point function matrix and, in the process of inversion, 
the transverse and
longitudinal components of the vector boson fields decouple. Having considered
the transverse degrees of freedom
in the preceding section, we can now limit ourselves to
the system formed by the longitudinal components of the photon
and of the $Z^0$ and by the Higgs  and the neutral Goldstone bosons, which we
denote by $S=\{A_\smallL,Z_\smallL,G_0,H\}$.
The two point functions  involving one vector boson  and one  scalar
are defined  extracting   $q^\mu$. In this way, $\g^S$ is the 4$\times$4
matrix of the two-point functions of $S$. 

The system $S$ includes unphysical degrees of freedom.
As we have noted in the introduction, even at the tree level the Green
functions of unphysical fields are modified by the choice to use 
the reduced generating functional $\g$ in place of the complete
functional ${\g^c}$ (see the App.\,A).
For the  purposes of this section, however,  the reduced functional simplifies
significantly the derivation without affecting the physical information we can
extract from $\g^S$.
In a way, this can be viewed as  a consequence of the fact 
that the cancellation between the unphysical degrees of  freedom
occurs independently of the gauge-fixing sector \cite{aoki,henri}.

Each row of $\g^S$ is connected by a STI. For instance, 
differentiating \equ{app3} with respect to $A^\mu$ and $c^\smalla$, we 
obtain for the first row 
\be
\left( c^2_\smallw -s_\smallw \g_{c_\smalla \gamma_3} \right) \g_{\smalla\smalla}^\smallL + 
\left( s_\smallw c_\smallw   + c_\smallw \g_{c_\smalla \gamma_3} \right) \g_{\smalla \smallz}^\smallL
+\g_{\smalla \smallG_0 }\g_{c_\smalla\gamma_0}  +  \g_{\smalla \smallh}
\g_{c_\smalla\gamma_\smallh}=0.
\ee
Similar identities can be derived for the other rows, so that 
the STI for the two-point functions can be written as 
$\g^S V_{c_\smalla}=0$, where $V_{c_\smalla}^\smallT
=(  c^2_\smallw  -s_\smallw \g_{c_\smalla\gamma_3},
s_\smallw c_\smallw   +  c_\smallw\g_{c_\smalla  
\gamma_3},\g_{c_\smalla\gamma_0},\g_{c_\smalla\gamma_\smallh})$.
Since $\phi$ includes the unphysical components of the photon and $Z^0$ fields
and since  we have eliminated the gauge fixing sector of the tree level
Lagrangian  in using the reduced functional (see \equ{redfun}),
it is perhaps  not surprising that there is no propagator for  $A_\smallL$
and $Z_\smallL$ and that $\det \g^S=0$  or  the  
rank of $\g^S$ is less than 4. 
In fact, $\g^S$ has another  linearly independent eigenvector $ V_{c_\smallz}$
with zero eigenvalue, corresponding to the set of STI 
obtained by differentiation wrt $c_\smallz$. 
Therefore,  the rank of $\g^S$ is at most 2 and that we
cannot use \equ{detgen} at this stage.
Moreover, the sub-matrix of $\g^S $ identified by the indices $G_0$ and $H$ has
the same rank as $\g^S$. 
This can be seen by noting that
 $V_{c_\smallz}$ and $V_{c_\smalla}$ can be orthogonalized in the
subspace  of the   $A^\smallL$ and $Z^\smallL$ components because
\be
\det\left(\begin{array}{ll}
c^2_\smallw - s_\smallw\g_{ c_\smalla\gamma_3} & 
s_\smallw c_\smallw   - s_\smallw\g_{c_\smallz \gamma_3}\\
s_\smallw c_\smallw + c_\smallw \g_{c_\smalla\gamma_3} & 
s^2_\smallw + c_\smallw \g_{c_\smallz \gamma_3}
\end{array}\right) =1 + O(\hbar)\neq 0, 
\ee
Having eliminated 
the unphysical longitudinal components of the vector bosons,
we can now concentrate on the  sub-matrix
\be
\g^{\cal H}=\left(\begin{array}{ll}
\g_{G_0 G_0} & \g_{G_0 H}\\
\g_{H G_0} & \g_{HH}\\
\end{array}\right),
\ee
whose rank is equal to the one of $\g^S$. Indeed, at arbitrary $q^2$, its
rank is 2, so that  \equ{detgen} is satisfied. $\g^{\cal H}$ is
very similar to the $\gamma-Z^0$ transverse mixing matrix.
Even if the CP violation mixes up physical and unphysical scalar
 fields at high perturbative orders, 
it is not difficult  to disentangle them taking advantage of the STI.
At $q^2=0$ the two STI obtained by differentiating wrt 
$c_{\smalla,\smallz} $ and $G_0$ imply that $\det \g^{\cal H}(0)=0$. 
This zero  is related to the $G^0$ field and is located at $q^2=0$ 
(in the standard $R_\xi$ gauge it would be at $q^2=\xi_\smallz \mz^2$) as a
consequence of the use of the reduced functional.
The remaining zero, at $q^2=\bar{s}_\smallh$, corresponds instead
to the physical pole of the Higgs boson and its location in the complex plane
is therefore gauge-independent, as it follows from \equ{detgen}. 
A discussion of the relation between the pole mass and the
conventionally  renormalized mass of the Higgs boson
 in this case can be found in \efe{si-kniehl}.

\section{Fermions}
\label{fermions}
The treatment of the fermionic sector is only slightly more involved than
that of the scalar sector. Again, we consider the most general case of mixing
and call $\g^f$ the matrix of the fermionic two-point functions,
$\g_{\bar{f}f'}$.
In the case of  massless neutrinos, there is no mixing in the leptonic sector 
and $\g^{lept}$ is  a diagonal matrix.
As a first step, we need to decompose  $\g^f$ into scalar pieces:
\be
\label{f1}
\g^f(p) = 
\Sigma_{\smallL}(p^2)  \not\!p \,P_{\smallL}+ 
\Sigma_{\smallR} (p^2) \not\!p \,P_{\smallR} +
\Sigma_{\smallD}(p^2) P_{\smallL} + 
\Sigma_{\smallD}^\dagger (p^2) P_{\smallR} 
\ee
where  $P_{\smallL,\smallR}=
\frac1{2} (1\mp \gamma_5)$ are
the left and right-handed  projectors. As can be seen by inverting $\g^f$,
the relevant quantity for 
the  fermion propagator matrix is  the matrix \cite{donoghue}
\be
\label{Kf} 
{ K_f}(p^2) = 
p^2 \,\Sigma_{\smallL} -  
\Sigma_{\smallD}^\dagger\,\, \Sigma_{\smallR}^{-1}\,
\Sigma_{\smallD} ,
\ee
where we have dropped the $p^2$ dependence of the $\Sigma$ matrices.
Since the determinant of this matrix appears in the denominator of the
fermion propagators, we want to study its zeros, i.e.\ the zeros of the
eigenvalues of $K_f$.
We recall that by  pseudo-hermiticity $\g^f
= \gamma^{\dagger}_{0}\, {\g}^{f\dagger}\,\gamma_{0}$, so that 
$\Sigma_{\smallL}^{\dagger}(p^2)=\Sigma_{\smallL}(p^2)$ and 
$ \Sigma_{\smallR}^{ \dagger}(p^2)=\Sigma_{\smallR}(p^2)$
(this is actually true below thresholds, but it 
does not affect our conclusions). Hence,  the matrix ${K}_f(p^2)$ 
is hermitian and can be diagonalized by means of 
a unitary transformation. Under the usual assumption
$\beta_i^\xi=0$,
the gauge-parameter dependence of $\g^f$ is described
by a Nielsen identity which has exactly the same form of \equ{Ngen}.
Setting furthermore $\rho^\xi=\gamma^\xi_\varphi=0$ for ease of notation
(the results would not change),
we have   $\Lambda=-\g_{\chi\bar{f}\eta_{f'}}$ and 
$\Lambda'=-\g_{\chi\bar{\eta}_f f'}$, which  have a Dirac structure and undergo
a decomposition analogous to \equ{f1}. Again by pseudo-hermiticity, 
we find that in this case $\Lambda_{\smallL,\smallR}=
(\Lambda'_{\smallL,\smallR})^\dagger$ and $\Lambda_\smallD=\Lambda'_\smallD$.
It is then straightforward  to verify that the 
components of $\g^f$ satisfy
\bea\label{f3} 
\de_\xi \Sigma_{\smallL} & = & {\Lambda}_{\smallL} \Sigma_{\smallD} +
\Sigma_{\smallL} {\Lambda}_{\smallD} +
{\Lambda}_{\smallD}^\dagger \Sigma_{\smallL} +
\Sigma_{\smallD}^\dagger {\Lambda}_{\smallL}^\dagger 
 \non\\
\de_\xi \Sigma_{\smallR} & = & {\Lambda}_{\smallD} \Sigma_{\smallR} +
\Sigma_{\smallR} {\Lambda}_{\smallD}^{ \dagger}
+{\Lambda}_{\smallR} \Sigma_{\smallD}^\dagger +
\Sigma_{\smallD} {\Lambda}_{\smallR}^{ \dagger} \\
\de_\xi \Sigma_{\smallD} & = & p^2 ({\Lambda}_{\smallR} \Sigma_{\smallL} +
\Sigma_{\smallR} {\Lambda}_{\smallL}^{ \dagger})+
\Lambda_\smallD \Sigma_\smallD + \Sigma_\smallD \Lambda_\smallD\ ,\non
\eea
from which it follows that
\be
\de_\xi  { K_f} = {K_f} \, F+  F^{ \dagger}\, K_f\, ,
\label{kappa}
\ee
with $F=\Lambda_\smallD - \Lambda_\smallR^\dagger \Sigma_\smallR^{-1}
\Sigma_\smallD$. Without using pseudo-hermiticity, we would have  $F'\neq
F^\dagger$ in place of $F^\dagger$ in the previous equation. \equ{kappa}
is in the form of \equ{Ngen} and therefore
${\cal D}_f\equiv \det K_f$ satisfies \equ{detgen}. We have therefore
algebraically reduced the problem in the fermionic case to the scalar one. 
In the case of mixing between $n$ fermions, 
the gauge-parameter independence of $n$ complex poles is thus warranted.
Again, this result holds for any choice of the 
fermion wave function renormalization and  relies solely on the $\beta^\xi_i=0$
assumption.

 The above proof  is new and valid in the full SM. For what concerns pure 
QED and QCD, the result that the pole masses of the 
electron and of the quark are gauge-independent 
is not new and has been obtained both using the Nielsen
identities \cite{bls} and in different ways \cite{kronfeld,poleqed}. 
In QED  (QCD) the situation simplifies considerably: writing 
$\g_{\bar{f}f} = B \not\! p - m A$, where $m$ is the mass of the electron
(quark), and decomposing $\Lambda$ in an analogous way, we find
\be
-\de_\xi A=\frac{p^2}{m} \,B \,\Lambda_\smallB + m \,A  \,\Lambda_\smalla; \ \ \ \ 
-\de_\xi B={m}\left( A  \,\Lambda_\smallB +  B \, \Lambda_\smalla\right), 
\ee
which could be tested up to $O(\alpha_s^2)$ against the general $R_\xi$ gauge
calculation of \efe{tarasov}. 

The proof of the IR finiteness of the fermions in the SM follows 
\efe{kronfeld} and the final discussion in Sec.\,3 and is already present 
{\it in nuce} in \efe{psw}.
For completeness, in App.\,B  we present the explicit gauge-parameter 
dependence of the one-loop fermionic self-energies in a general $R_\xi$ gauge
for the full SM.
Remembering that $\Lambda$ first occur at the one-loop level,
it is straightforward to see that 
they satisfy  the Nielsen identities \equ{f3}.
This completes the set of expressions given in \efe{ds92} and is very useful 
in particular applications. For instance, Eqs.\,(\ref{sigmas}-\ref{sigmar})
 have been used in \efe{ckm}
to discuss the gauge dependence of the one-loop definition of the CKM matrix. 
Indeed, as noted in the previous section, the renormalization of the mixing
parameters is a delicate subject for what concerns the gauge-parameter
dependence. An adequate framework for studying it is the Background Field 
Method \cite{BFM}.
In the case of the fermion mixing a comprehensive analysis
has been presented in \efe{ckm}. 

\section{Application to physical amplitudes} 
\label{ampl}
In this section we apply the formalism of the Nielsen identities to 
four-fermion physical amplitudes and study the mechanism of gauge 
cancellations at any order in perturbation theory. 
Our purpose here is not to prove the gauge-independence of the physical 
amplitudes, a result which was accomplished in full generality long ago 
at the level of the generating functional \cite{brs}. We would rather like to 
study a specific example and carry out the analysis at an arbitrary order in
perturbation theory. The use of the Nielsen identities allows 
us to uncover the regularities 
of the gauge recombinations between the different components
(vertices, boxes and self-energies) in great generality.
The following derivation is formally independent
of the perturbative expansion of the Green functions. In other words, 
if we work at order $n$ in perturbation  theory 
the Green functions have to be expanded up to this order,  but 
the factorization works independently of that.
At the one-loop level, a similar  factorization is also accomplished 
diagrammatically 
by the Pinch Technique (PT) \cite{pinch}, whose extension at higher orders 
has however proved problematic. Unlike the PT, 
the Nielsen identities control only the gauge parameter
variation and cannot be used to construct explicitly gauge-independent
proper functions which satisfy basic requirements and  tree-level-like
Ward identities. However, they may prove useful in the search for the
higher-order extension of the PT.
The analysis of this section gives us also the opportunity to
present explicitly the  Nielsen identities for vertices and boxes
involving fermions, which are interesting in their own respect as they appear
in most phenomenological applications.

We first 
consider the truncated Green function  $Z^{trunc}_{\bar{I} J \bar{K} N}$ 
(see e.g.\ \cite{itz}) for a generic  four fermion process 
$f_{\bar{I}} f_J \longrightarrow  f_{\bar{K}} f_N$  
and we decompose it in terms of irreducible diagrams and propagators. 
We will  use  capital and lowercase letters to denote fermions and bosonic 
fields (scalar as well as gauge vector bosons), respectively.
Therefore,  $Z^c_{\bar{I} J}$ and $ Z^c_{i j}$ 
are  the propagator functions  of  fermions and  bosons. 
Following the convention of the preceding sections, 
irreducible boxes and vertices are denoted by
 $\g_{ \bar{I} J \bar{K}N}, \g_{ \bar{I}J i}$, and $ \g_{j \bar{K}N}$. 
To keep the notation simple, we drop Lorentz indices and the dependence on the
external momenta. 
The physical amplitude ${\cal M}_{\bar{I} J \bar{K} N}$ for our process 
is obtained from $Z^{trunc}_{\bar{I} J \bar{K} N}$ using  the LSZ 
reduction formula \cite{itz}, 
which in the case of fermion with mixing reads \cite{aoki}
\begin{eqnarray}
  \label{am_7} {\cal M}_{\bar{I} J \bar{K} N}=
\lim_{on-shell}  
{\wt{Z}}^{1/2}_{\bar{I} I'}\, 
{\wt{Z}}^{1/2}_{\bar{J}' J}\,\, 
Z^{trunc}_{\bar{I}' J' \bar{K}' N'} \,\, 
{\wt{Z}}^{1/2}_{\bar{K} K'} \, 
{\wt{Z}}^{1/2}_{\bar{N}' N} ,
\end{eqnarray}
where the on-shell limit includes the projection on the asymptotic states and
 $\wt{Z}$ controls the relation between the asymptotic
states and the renormalized  spinors:
\be
{\wt{Z}}^{1/2}_{\bar{I}J}\, u_{as,J}= u_I.
\label{asym}
\ee
The matrix ${\wt{Z}}$ can be computed from  the conditions \cite{aoki}
(quantum equations of motion)
\bea 
\g_{\bar{I}J} \,u_{J}(m_J)=0;  \ \ &&\ \ \
\bar{u}_{I}(m_I) \,\g_{\bar{I}J} =0\non\\  
 \frac{\tilde{Z}_{II}}{\not\!p -m_I}
 \,\g_{\bar{I}I} \,u_{I}(m_I)=u_{I}(m_I);\ &&\ \ 
\bar{u}_{I}(m_I)\,\g_{\bar{I}I}\,\frac{\tilde{Z}_{II}}
{\not\!p -m_I}  =\bar{u}_{I}(m_I),
\label{qem}
\eea
using the fact that $(\not \!p -m_J) \,u_{as}(m_J)=0$ at any order 
by definition.
Of course, $\wt{Z}$ should be decomposed in left and
right-handed parts, $\wt{Z}=\wt{Z}^\smallL P_\smallL +
\wt{Z}^\smallR P_\smallR$. 
 Notice that the first line of \eqs{qem} implies $\det K_f =0$ and
consequently includes  the
requirement that the mass parameters of the external fermions
are renormalized on the poles of the propagators (see
Sec.\,\ref{fermions}).
 Strictly speaking, the LSZ formalism applies only to
stable external states, i.e.\  to the electron and  neutrinos
and, to a good approximation,  to the muon. Nevertheless, we will 
consider here the general case of mixing.
We also stress that the LSZ factors $\wt{Z}$ should not  be confused
with the 
wave-function renormalization factors for the external fields. Of course, the
latter can be {\it chosen} by imposing \eqs{qem} together with
$\wt{Z}={\bf 1}$ ({\em on-shell}
scheme \cite{aoki}), but there is in general no restriction on their choice
(see also \cite{ckm}) and it is even possible to avoid them altogether, in
which case $\wt{Z}$ is divergent.
Once the wave-function renormalization has been defined, for instance
through a minimal subtraction, the  factors $\wt{Z}$ can be computed
from \eqs{qem}.

As a first step, 
we consider the gauge variation of the truncated Green function
$Z^{trunc}_{\bar{I} J \bar{K} N}$
In the most general case of mixing,  $Z^{trunc}_{\bar{I} J \bar{K} N}$ 
is decomposed in the following  blocks (we sum over repeated indices) 
\begin{eqnarray}
  \label{am_1}
Z^{trunc}_{\bar{I} J \bar{K} N} &=& 
i\,\g_{\bar{I} J  \bar{K} N} - \g_{\bar{I}J i} \ Z^c_{i j}\ \g_{j \bar{K} N} - 
\g_{ \bar{I}N i} \ Z^c_{i j}\ \g_{j \bar{K} J},
\end{eqnarray}
from which we obtain
\begin{eqnarray}
  \label{am_2}
\de_\xi Z^{trunc}_{\bar{I} J \bar{K} N} 
&=& 
i\,\de_\xi \g_{ \bar{I} J  \bar{K}N}  
\nonumber \\
&-& 
  \left(\de_\xi \g_{ \bar{I} Ji}\right)   Z^c_{i j}\ \g_{j \bar{K} N} - 
  \g_{ \bar{I} Ji}  \left(\de_\xi Z^c_{i j} \right) \g_{j \bar{K} N} - 
  \g_{\bar{I}J i}  \ Z^c_{i j} \ \de_\xi  \g_{j \bar{K} N}  
 \\
&-& 
  \left(\de_\xi \g_{ \bar{I}N i}\right)  Z^c_{i j} \ \g_{j \bar{K} J} -
  \g_{ \bar{I}N i} \left( \de_\xi Z^c_{i j} \right)\g_{j \bar{K} J} -
  \g_{ \bar{I}N i}  \ Z^c_{i j}\  \de_\xi \g_{j \bar{K} J} \,.\nonumber
\end{eqnarray}
We can compute the different contributions $\de_\xi \g_{J \bar{I} N \bar{K}}, 
 \de_\xi \g_{J \bar{I} i}$, and 
$ \de_\xi Z^c_{i j}$ using the Nielsen identities. 
The identity for the propagator functions $ Z^c_{i j}$ and $Z^c_{\bar{I} J}$ 
is easily  derived from the identity for the 
irreducible {\bf two-point functions} $\g_{i j}$ and $\g_{\bar{I} J}$. 
As we have seen, the general form of the latter is 
\be
\de_\xi \g_{ab}= -
\g_{ac}\, \g_{\chi b \gamma_c}-\g_{\chi a\gamma_c}\,\g_{cb},
\label{ni2}
\ee
where the indices $a,b$ apply to both the bosonic and fermionic case. 
As usual,  we employ the procedure of Sec.\,2, remove all  tadpoles,
set $\delta_t=\delta_t^{CP}=0$, and assume   $\beta^\xi_i=0,~\forall i$
(this is consistent with the LSZ use of pole masses, as we have seen). 
Concerning the $\rho^\xi$ and
$\gamma^\xi_\varphi$ factors, we avoid them here in order to keep 
the formulas simple.
\begin{figure}[t]
\centerline{
\psfig{figure=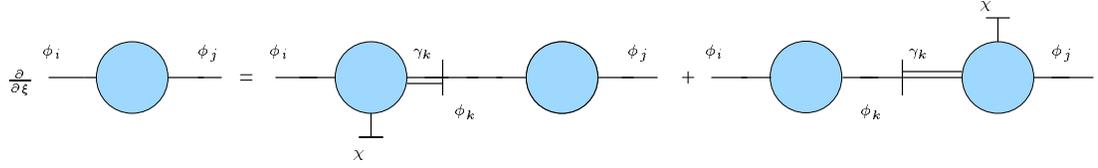,rheight=.6in}  }
\caption{\sf Nielsen identity for the two-point function $\g_{\phi_i\phi_j}$.}
\label{fig2p}
\end{figure}
However, following the discussion in Sec.\,2, they are bound to drop out of 
the amplitude and this can be explicitly verified.
\equ{ni2} can be graphically represented in the very simple way shown in 
Fig.\,\ref{fig2p}. 
Notice that the momentum  flows along the horizontal line
and that the insertion of the static source $\chi$  does not carry momentum,
unlike the one of $\gamma_\varphi$.

Using  the relations $Z^c_{i j} \,\g_{j k} =i \delta_{i k}$ and 
$Z^c_{\bar{I} K} \, \g_{\bar{K} J} = i \delta_{I J} {\bf 1}$, where 
${\bf 1}$ is the identity matrix for the Dirac indices, we obtain 
the Nielsen identities for the {\bf propagator functions}, which read
\begin{eqnarray}
 \de_\xi Z^c_{i j} &=&  Z^c_{i k}\, \g_{\chi k \gamma_j }   +
\g_{\chi \gamma_i k} \,Z^c_{k j} ,\\ 
\de_\xi Z^c_{\bar{I} J} &=& 
Z^c_{\bar{I} K} \g_{\chi \bar{K}  {\eta}_J } +
\g_{\chi \bar{\eta}_I K }  Z^c_{\bar{K} J},   \label{am_3}
\end{eqnarray}
for bosons and fermions, respectively. Graphically, these identities can be 
represented by Fig.\,\ref{fig2p} after replacing the blobs with the $\chi$
insertion by their mirror images and exchanging the corresponding indices.
For the {\bf three-point functions} we have 
\begin{eqnarray}
  \label{am_4}
-\de_\xi \g_{\bar{I} J i} 
&=& 
\g_{\chi \gamma_m  \bar{I} J} \g_{m i} +  
       \g_{\chi i \gamma_m }  \g_{m \bar{I} J} 
\nonumber \\ &+& 
\g_{\bar{I} K i} \g_{\chi \bar{\eta}_K J } + 
       \g_{\bar{I} K} \g_{\chi i \bar{\eta}_K J } + 
       \g_{\chi   \bar{I}  {\eta}_K }  \g_{\bar{K} J i} + 
       \g_{\chi   i \bar{I}  {\eta}_K }  \g_{\bar{K} J}%
\,. 
\end{eqnarray}
We see that the gauge-dependent terms of the form of $\g_{\chi \gamma_i j}$
introduced by the propagators in \equ{am_2}
are exactly cancelled by the last term in the first line of \equ{am_4}, i.e.\
 by the vertices alone. Therefore, the boxes are not necessary to remove the 
gauge-dependence of the internal self-energies. 
The identity for the {\bf four-point functions} is 
\begin{eqnarray}
  \label{am_5}
-\de_\xi \g_{\bar{I} J \bar{K} N} 
&=& 
 \g_{\bar{I} J m}   \, \g_{\chi \gamma_m  \bar{K} N}  +
 \g_{\bar{I} N m}   \, \g_{\chi \gamma_m  \bar{K} J} + 
 \g_{\chi \gamma_m  \bar{I} J} \,\g_{m  \bar{K} N} +  
 \g_{\chi \gamma_m  \bar{I} N} \,\g_{m  \bar{K} J}  
\nonumber 
\\ &+& 
 \g_{\bar{I} S} \,\g_{\chi \bar{\eta}_S J \bar{K} N} + 
          \g_{\bar{K} S} \,\g_{\chi  \bar{I}J \bar{\eta}_S  N} + 
          \g_{\chi \bar{I} J \bar{K} {\eta}_S}\, \g_{\bar{S} N} + 
          \g_{\chi \bar{I} {\eta}_S \bar{K}N } \,\g_{\bar{S} J}
\\ &+&
 \g_{\bar{I} J \bar{K} S}\, \g_{\chi \bar{\eta}_S N} + 
          \g_{\bar{I} S \bar{K} N} \,\g_{\chi \bar{\eta}_S J} + 
          \g_{\chi \bar{I} {\eta}_S} \,\g_{\bar{S} J \bar{K} N} + 
          \g_{\chi \bar{K} {\eta}_S} \,\g_{\bar{I} J \bar{S} N} \,. 
\nonumber 
\end{eqnarray}
We now distinguish between the different Green functions containing
the source $\chi$:
\begin{enumerate}
\item  Terms of the form $\g_{\chi \gamma_i \bar{I} J}$ are present both in 
the gauge variation of the boxes (first line) 
and in the one of the vertices (first term). They 
cancel against each other in the sum  (\ref{am_2}) according to the pattern
\begin{eqnarray}
  \label{am_6}
&&  \underbrace{ \g_{\chi \gamma_m  \bar{I} J}\ \g_{m i} \ Z^c_{i j}\ 
    \g_{j \bar{K} N}}_{-(\de_\xi \g_{ \bar{I}j i}) \ Z^c_{i j}\ 
\g_{j \bar{K} N}} -
  \underbrace{i\,\g_{\chi \gamma_m  \bar{I} J}\ \g_{m  \bar{K} N}}_{i\,\de_\xi 
\g_{\bar{I}J  \bar{K}N}} = 0 \ ,\non
\end{eqnarray}
where we have specified which part of  \equ{am_2} generates each term. 

\item  The factors 
containing $\g_{\chi i\bar{\eta}_K J}$ in the second line of 
\equ{am_4} and  $\g_{\chi \bar{\eta}_S J \bar{K} N}$ (the whole second line 
of \equ{am_5}) always multiply a two-point function of the external fermions
like $\g_{\bar{I}J}$.
When they are contracted with the external spinors, these terms vanish, 
as a consequence of \equ{qem}.

\item  The remaining terms contain Green functions of the kind 
$\g_{\chi \bar{\eta}_I J}$ and $\g_{\chi \bar{I}\eta_J} $
which multiply vertices and boxes in \eqs{am_4} and (\ref{am_5}),
respectively. As we will see in a moment, 
they are cancelled by the LSZ factors.
\end{enumerate}
Adding together the various pieces,  the gauge-parameter variation
of the on-shell 
truncated Green function can be expressed in terms of the truncated
function itself:
\be
-\de_\xi  Z^{trunc}_{\bar{I} J \bar{K} N}|_{on-shell}=
\g_{\chi \bar{I} \eta_S} \,Z^{trunc}_{\bar{S} J \bar{K} N} 
+\g_{\chi \bar{K} \eta_S} \,Z^{trunc}_{\bar{I} J \bar{S} N }
+Z^{trunc}_{\bar{I} S \bar{K} N }\, \g_{\chi \bar{\eta}_S J}
+Z^{trunc}_{\bar{I} J \bar{K} S }\, \g_{\chi \bar{\eta}_S N},
\ee
according to the usual form for the Nielsen identities. Of course, this 
on-shell factorization
holds in general  for any amputated Green function, as it follows from the 
gauge independence of the $S$-matrix.

We are now ready to apply the LSZ reduction formula. 
The gauge variation of the factor $\wt{Z}$  can be computed from 
\equ{asym} and \equ{qem} using the Nielsen identities for the two-point 
functions and the gauge-independence of the asymptotic spinors $u_{as, I}$.
We then obtain
\be
\lim_{on-shell} \de_\xi {\wt{Z}}^{1/2}_{\bar{I}J} \, u_{as,J}=
\g_{\chi\bar{I} \eta_S} {\wt{Z}}^{1/2}_{\bar{S}J}\, u_{as,J},
\ee
where $\g_{\chi\bar{I} \eta_S}$ is calculated on-shell, 
from which the final cancellation of the gauge-dependence follows.

If some of the $\beta^\xi_i$ do not vanish,  the cancellations 
do not operate any longer 
and the amplitude turns out to be gauge parameter dependent \cite{brs}.  
An explicit example has been considered in \cite{ckm}, for the $W$ decay 
into quarks: if the CKM counterterm is gauge-dependent,
 the amplitude  depends on the gauge parameters too. 
On the other hand, the above proof relies neither on a specific choice of 
renormalization of the unphysical parameters, nor on the regularization 
scheme adopted (provided  the STI  have been restored 
order by order).

\section{Summary}
\label{summary}
We have  introduced   the Nielsen identities of the SM and used the
problem of the definition of mass as a demonstrative example. In this context
we have obtained some new results: we have proven to all orders in perturbation
theory the gauge-parameter independence of the 
complex pole associated to any physical  particle of the SM.
We have considered the cases of the vector bosons, scalars and
fermions in great generality, allowing for arbitrary mixing patterns.
Particular attention has been paid to the case of the $W$ boson, which is 
simpler because of the absence of mixing and has been chosen to
 illustrate some features  common to all cases.
Most of the proofs hold without modifications also in some  extensions
of the SM, like non-supersymmetric two-Higgs-doublet models.

We have derived  identities for the gauge-dependence 
 of all the two-point functions of the SM, 
both for bosons and fermions, as well as for vertices and boxes 
involving external fermions. 
  Using these expressions, we have shown the 
explicit mechanism of gauge cancellations which leads to gauge-independent 
four fermion amplitudes to all orders in 
the most general case of fermion and boson mixings and of CP violation.
The formalism introduced in this paper, supplemented by the material
given in App.\,A (the Lagrangian involving the BRST sources), should allow 
for a very simple derivation of the Nielsen identities for any proper Green 
function  in  the electroweak SM and in QCD. 

We have also extensively discussed the renormalization of the Nielsen 
identities with  an arbitrary regularization,
in  the case the Nielsen identities (but not the STI) are broken
by renormalization. 
In that case the identities are deformed by new terms, which we have 
identified in full  generality  and computed in a few cases of particular 
interest. We have also derived  new results  concerning the 
infrared-finiteness of the $W$ pole mass and the photon two-point function 
at $q^2=0$ in the SM. 
For completeness, we  report in App.\,C the expressions for the fermionic
one-loop self-energies in a generic $R_\xi$ gauge.

In conclusion, the formalism of the Nielsen identities 
can  be  useful  in various applications: (i)  at the  conceptual  level, 
for the identification of gauge-independent quantities such as 
invariant charges \cite{kraus-si-h} and for the gauge-independent
 definition of renormalized parameters \cite{ckm}; 
(ii) at the  practical level, because  in higher orders calculations it 
is generally simpler to compute the gauge-dependence using the Nielsen
identities, and because these identities allow for powerful  checks. 
It deserves to be better known to  theorists. 

\subsection*{Acknowledgments}
We are grateful to C. Becchi, M. Passera, A. Sirlin,
 and W. Zimmermann for interesting discussions
and to M. Steinhauser for useful communications  
and  a careful reading of the manuscript. 
This work has been supported in part 
by the Bundesministerium f{\"u}r Bildung  und
Forschung under contract 06 TM 874 and by the DFG project Li 519/2-2.

%%%%%%%%%%%%%%%%%%%%%%%%%%%%%%%%%%%%%%%%%%%%%%%
\begin{appendletterA}
\section*{A. Nielsen identities for pedestrians}
 
The aim of this Appendix is to  
review very briefly the formalism of Slavnov-Taylor Identities
(STI) in the case of the Nielsen identities 
and to provide some  material necessary for the explicit calculation of 
the Green functions involving the BRST sources.
For a non-expert introduction
to the STI for specific physical amplitudes, we refer to \cite{ghs}.
First, we recall that in our conventions the gauge-fixing term in the 
SM Lagrangian is given by 
\begin{eqnarray} 
\hspace{-2cm} {\cal{L}}_{GF}  
        & = &   
     -\frac1{2\xi_\smalla} \left(\partial^{\mu}A_{\mu}\right)^2
-\frac1{2\xi_\smallz^{(1)}}\left(
               \partial^{\mu}Z_{\mu} -     
                            {{\xi_\smallz^{(2)}\,{\sqrt{{{g'}^2} + {{g}^2}}}}\over 2}     
                           \, v \,{G^{0}} \right)^2 \non
 \\          & - &  \frac1{\xi_\smallw^{(1)}} \left|
               \partial^{\mu}W^+_{\mu} -     
             \frac{i \xi_\smallw^{(2)} \, g }{2}  \, v \, {G^+} \right|^2  
        -\frac1{2\xi_g}  \left( \partial^{\mu} G^{b}_{\mu} \right)^2  .
\end{eqnarray}  
We always set $\xi_{\smallw,\smallz}\equiv\xi_{\smallw,\smallz}^{(1)}=
\xi_{\smallw,\smallz}^{(2)}$,
i.e.\ we confine  ourselves to the restricted 't Hooft gauge.
Our starting point is the complete generating vertex functional $\g^{c}$, 
which  generates the one-particle-irreducible Green functions.
In order to simplify the structure of the STI, 
it is convenient to introduce for linear gauge-fixings 
 a {\it reduced} generating functional $\g$ (sometimes indicated by $\gh$ in
 the literature),
which differs from ${\g^c}$ by a local term, corresponding to the
gauge-fixing part of the Lagrangian:
\be
\g= {\g^c} - \int d^4 x \ {\cal{L}}_{GF}.
\label{redfun}
\ee
In practice, the STI obtained from $\g$ coincide with the STI obtained from 
${\g^c}$ after  implementation of the ghost equation \cite{itz}.
Of course, one should keep in mind that 
the Green functions involving unphysical fields 
generated by $\g$ coincide with the ones 
generated by ${\g^c}$ only up to constant  terms. For example,
one has 
$\g_{W_\mu W_\nu}^{(0)}= {\g}_{W_\mu W_\nu}^{c(0)} +
p^\mu p^\nu/\xi_\smallw$ and
$\g_{G^+ G^-}^{(0)}= {\g}_{G^+ G^-}^{c(0)}+\xi_\smallw \mw^2$
at the tree level,
while the difference at higher orders depends only on the renormalization of
the $W$ field and of the gauge parameters. As we have eliminated the 
classical gauge-fixing, it is clear that     
$\g_{\smallw^+ G^-}\neq 0$ already at tree level.

The invariance of the action under BRST transformations implies 
the STI for the functional $\g$ (see for ex. \cite{itz}),
\be
{{\cal S}\left( \Gamma \right) }= \int d^4 x \,\left[ \partial_\mu c^0 
\frac{\del\g}{\del B_\mu} + 
\sum_\varphi \, 
\frac{\del\g}{\del \gam_\varphi} \,\,\frac{\del \g}{\del \varphi} 
\right]  =0, 
\label{app3}
\ee
where $B_\mu$ and $c^0$ are the gauge boson for the $U(1)$ abelian 
factor of the gauge group and the corresponding ghost.
$\varphi$ stands for any other quantum field in the SM Lagrangian
(gauge fields, scalars, ghosts, and fermions) 
and $\gam_\varphi$ is  the   BRST source associated to $\varphi$  
and is coupled to the {\it non-linear} BRST variation of $\varphi$   
in the classical action. 
In the case of a  fermion $f_I$ the spinorial source is denoted by $\eta_I$.  
We also introduce the Slavnov-Taylor operator ${{\cal S}_\Gamma}$ defined by
\be
{{\cal S}_\Gamma}= \int d^4 x \, \left\{  
\partial_\mu c^0 \,
\frac{\del}{\del B_\mu} +  
\sum_\varphi \, \left[
\frac{\del\g}{\del \gam_\varphi} \,\,\frac{\del}{\del \varphi}+
\frac{\del\g}{\del \varphi} \,\,\frac{\del}{\del \gam_\varphi}\right] \right\};
%\ \ \ \ {{\cal S}_\Gamma}\,\g=0.
\label{app3.1}
\ee
By functional differentiation of \equ{app3} with respect to 
some SM fields one gets the Slavnov-Taylor 
Identities (STI). Electric and ghost charge conservation, as well as 
Lorentz invariance, should be taken into account, according to the examples
given in the text.

In order to obtain the Nielsen identities for the gauge parameter 
dependence of irreducible Green functions, we have to consider the case 
of extended BRST symmetry \cite{zuber}, which involves also the
transformation of the gauge parameters; \equ{app3} takes then the form
\be
{{\cal S}\left( \Gamma \right)} \,  + \sum_i \,\chi_i  \,\de_{\xi_i} \g=0, 
\ee
from which \equ{nielsen} follows  after differentiating wrt $\chi$ and setting 
$\chi=0$. 
In the fermionic sector the expressions are slightly complicated by the 
anticommutation relations   and the Nielsen identity becomes
\be
\de_\xi \g^{fer}= \sum_I
\left[ \frac{\g\stackrel{\leftarrow}{\del}}{\del \psi_I} \,\,
\frac{\stackrel{\rightarrow}{\del} \de_\chi \g}{\del \bar{\eta}_I}- 
\frac{\de_\chi \g\stackrel{\leftarrow}{\del} }{\del \psi_I}\,\,
\frac{\stackrel{\rightarrow}{\del}\g}{\del \bar{\eta}_I} 
+ (\psi_I\leftrightarrow \eta_I)\right],
\ee
where $\de_\chi= \de/ \de\chi$ and the arrows indicate the direction in which 
the functional derivative wrt the fermionic field acts (this is important for
anticommuting fields).

We have seen that both the Nielsen identities and the STI contain Green
functions involving the BRST sources $\gamma_\varphi$ and $\eta_f$ (for fermions)
associated to the various fields of the SM. If we want to compute these
Green functions at a given order in perturbation theory, we need to know how
the sources are coupled to the fields. 
To this end, we  give below the complete action 
involving the BRST sources, which can be useful as a reference and to obtain
the Feynman rules necessary for actual calculations
involving $\gamma_\varphi$ and $\eta_f$. Apart from the well-known 
Feynman rules of the SM (see for instance the second paper in \cite{hollik}),
nothing else is needed to evaluate the unconventional objects that appear in
the identities.
Using the convention $Z_\mu = c_\smallw W^3_\mu + s_\smallw B_\mu$, where 
$W^3_\mu, B_\mu$ are the third component of the triplet of $SU(2)_L$ and 
the $U_Y(1)$ gauge boson, respectively,  we have 
\bea \label{app4}
 {\cal{L}}_{BRST}  
        & = &   
         \gam^{\mu}_3 \left\{ c_\smallw \de_\mu c^\smallz -
 s_\smallw \de_{\mu} c^{\smalla} - i g \left[  W^{+}_{\mu}  c^{-} - 
                W^{-}_{\mu}  c^{+} \right]     \right\}   
\non\\      & + &   
          \gam^{\mp\mu}_\smallw  \left\{\!\de_{\mu} c^{\pm} \mp i e 
          W^{\pm}_{\mu}  \left( c^{\smalla} -       
          \frac{c_{\smallw}}{s_{\smallw}} c^{\smallz} \right) \pm i e c^{\pm}\left[       
           A_{\mu}  - \!\frac{c_{\smallw}}{s_{\smallw}}      
           Z_{\mu} \right] \right\}  
\non\\      & + &   
          \gam^{a\mu} \left\{ \de_{\mu} c^{a} - g_{s} f^{abc}  
          G^{b}_{\mu} c^{c} \right\}  
-          \gamma_{c^3} \left\{  i g  c^{+} c^{-}   \right\}  
+ \gamma_{c^{\mp}} \left\{ \mp \frac{i e}{2} c^{\pm} \left( c^{\smalla} 
-        \frac{c_{\smallw}}{s_{\smallw}} c^{\smallz} \right)\right\} 
\non\\      & + &    
          \gamma_{c^{a}}  \left\{  \frac{g_s}{2}  f^{abc} c^{b} c^{c}\right\}  
+ \gam^{\smallh} \left\{  \frac{i  g}{2} \left[ 
                   G^{+}  c^{-} -         
                G^{-} c^{+} \right] +  
                \frac{g}{2  c_{\smallw}}       
                 G^{0}  c^{\smallz} \right\}   
\non\\      & + &   \gam^{\mp} \left\{ \pm \frac{i g}{2 }  
                \left[ H  + v \pm i       
                G^{0}  \right] c^{\pm}  
                \mp i e G^{\pm}    \left(c^{\smalla} -  
                \frac{c^{2}_{\smallw} - s^{2}_{\smallw}}{2 c_{\smallw} s_{\smallw}} c^{\smallz}      
                \right) \right\}   
\non\\      & + &   \gam^{0} \left\{  \frac{g}{2 } \left[G^{+} c^{-} +
                G^{-} c^{+} \right] -  
                \frac{g}{2  c_{\smallw}}       
                \left( H + v \right) c^{\smallz} \right\}  
\non\\      & + & i  \left( \b{\eta}_{\nu} ,  \b{\eta}^{L}_{l} \right)  
                     \left(  
                       \ber{c}   
                       \dms{\frac{g }{\sqrt{2}}}\,      
                       l^L \,c^{+} %
                      + \dms{\frac{g}{2}}\,\frac{c^{\smallz}}{c_\smallw} \nu  
                                   \vspace{.2cm}       \\  
                        \dms{\frac{g }{\sqrt{2}}}  
                        \nu c^{-} - e      
                        \left[ Q_{l} c^{\smalla}      
                          + \left(\dms{\frac{1}{2 s_{\smallw} }} +  
                            Q_{l} \dms{s_{\smallw}}      
                          \right)\frac{c^{\smallz}}{c_\smallw} \right] l^L   
                                             \eer  
                     \right)  
\vspace{.2cm} 
\non\\      & + &  i\left( \b{\eta}^{L}_{u} ,  \b{\eta}^{L}_{d} \right)  
                     \left(  
                       \ber{c}   
                       \dms{\frac{gV_{u\,d} }{\sqrt{2}}}      
                       d^L c^{+} - e \left[ Q_{u} c^{\smalla}      
                         - \left(\dms{\frac{1}{2 s_{\smallw} }} -  
                           Q_{u} \dms{s_{\smallw}}      
                         \right) \frac{c^{\smallz}}{c_\smallw} \right] u^L  
                       +  
                       g_s\frac{\lambda^a}{2} u^L  c_{a}   
                      \vspace{.2cm} \\  
                        \dms{\frac{g V^*_{u\,d}}{\sqrt{2}}}  
                        u^L c^{-} - e      
                        \left[ Q_{d} c^{\smalla}      
                          + \left(\dms{\frac{1}{2 s_{\smallw}}} +  
                            Q_{d} \dms{s_{\smallw}}      
                          \right) \frac{c^{\smallz}}{c_\smallw} \right] d^L
                        +   g_s\frac{\lambda^a}{2}  d^L  c_{a} 
                       \eer  
                     \right)  
\non\\      & - & i \,\b{\eta}^{R}_{l}  \left\{ e Q_{l} \left( c^{\smalla} +  
                             \dms{\frac{s_{\smallw}}{c_{\smallw}}}      
                             c^{\smallz} \right) l^{R}   \right\}  
+ i\,\b{\eta}^{R}_{u}   \left\{  -e \,Q_{u} \left( c^{\smalla} +  
                             \dms{\frac{s_{\smallw}}{c_{\smallw}}}      
                             c^{\smallz} \right) u^{R}   +    
                           g_s\frac{\lambda^a}{2}  u^{R}  c_{a}   
                         \right\} 
\non\\      & + &  i\,\b{\eta}^{R}_{d}  \left\{  -e \,Q_{d} \left( c^{\smalla} +  
                            \dms{\frac{s_{\smallw}}{c_{\smallw}}}      
                            c^{\smallz} \right) d^{R}   +   
                          g_s\frac{\lambda^a}{2} d^{R} c_{a}  \right\} + {\rm h.c.}
\eea 
where $\lambda^a$ are the Gell-Mann matrices, $R$ and $L$ indicate the 
right and left-handed components of the fermion fields,
 and  $s_\smallw=\sin \theta_\smallw$, 
$c_\smallw=\cos\theta_\smallw$. The hermitian conjugate for the fermionic
part is added at the end. The ghost charge of the various sources,
which is important in writing the STI, can be inferred by \equ{app4}, 
assigning a number +1 to the ghosts and requiring ${\cal{L}}$ to be ghost 
charge neutral.
No BRST source needs to be  introduced for the abelian vector field and for its
ghost. $\gamma^3_\mu$ is the source of the BRST transformation of the third 
component of the gauge boson triplet. 

The last ingredient for the calculation of the Green functions involving
the source $\chi$, characteristic of the Nielsen identities, are the couplings 
of $\chi$ with the other fields.
There is a  source $\chi_i$ associated to any  gauge parameter 
$\xi_i$ \footnote{Having set the two gauge parameters
$\xi_i^{(1,2)}$ equal to each other, we can work with only one source
$\chi_i$. This differs slightly from the procedure adopted in \cite{ckm}, 
where  two distinct sources $\chi_i^{(1,2)}$ were kept.}. The relevant
Lagrangian takes the form:
\bea
\label{chisources}
 {\cal{L}}_{\chi}&=& -\frac{\chi_g}{2\xi_g}\,\bc^{a} \,\de_\mu 
G^{a,\mu} 
-\frac{\chi_\smalla}{2\xi_\smalla} \,\bc^{\smalla} \,\de_\mu A^\mu  
-\frac{\chi_\smallz}{2\xi_\smallz} \,\bc^{\smallz} \left(\de_\mu Z^\mu +
\xi_\smallz \mz G_0 \right) \non\\
&-&\frac{\chi_\smallw}{2\xi_\smallw} \left[\bc^{\,+} \left(
\de^\mu W_\mu^- -i \xi_\smallw \mw G^- \right)+
\bc^{\,-} \left(
\de^\mu W_\mu^+ +i \xi_\smallw \mw G^+ \right)\right]
\eea 
\end{appendletterA}

\section*{B. Nielsen identities and regularization}
\begin{appendletterB}
In this appendix we clarify the meaning of \equ{nielsenren}
and show how its structure is preserved if the STI are enforced 
at each perturbative order by means of appropriate non-invariant counterterms.

Let us  consider a non-invariant regularization, such as dimensional
regularization in the implementation of \efe{dieter}, and proceed to 
impose the renormalization conditions according to the procedure outlined 
in Sec.\,\ref{sec2}. 
At a given order $n$ in perturbation theory the STI are violated.
We now assume that at order $n-1$ the STI have been restored by the
introduction of appropriate non-invariant counterterms.
Following the discussion of Sec.\,\ref{sec2},
the Nielsen identity  corresponding to the extended BRST symmetry
can be written, at order $n-1$, in the following form 
(here we consider explicitly different gauge-fixing parameters)
\be\label{app_2.1}
\left[ {\cal S}\left( \Gamma \right) +  \sum_i 
\chi \left( \partial_{\xi_i}\Gamma + \Delta_{\chi_i} \cdot \Gamma \right)
\right]^{(n-1)} = 0,
\ee
where
\bea\label{app_2.2}
&&\left( \Delta_{\chi_i} \cdot \Gamma \right)^{(n-1)} =
\sum_{m=1}^{n-1} \Delta^{(m)}_{\chi_i} \cdot \Gamma^{(n-m-1)} = \\
&&\sum_{m=1}^{n} \left(  \rho^{\xi,(m)}_{ij} \,\partial_{\xi_j} + 
\sum_{j} \beta_j^{\xi_i,(m)} \,\frac{\de}{\de p_j}  +
\sum_\varphi \gamma_\varphi^{\xi_i,(m)} \,{\cal N}_\varphi +
\delta^{\xi_i,(m)}_t \int d^4 x \frac{\delta}{\delta H(x)} \right)
\Gamma^{(n-m-1)}.
\nonumber
\eea
The matrices $\beta_j^{\xi_i,(m)},\gamma_\varphi^{\xi_i,(m)},
\delta^{\xi_i,(m)}_t$, and $\rho^{\xi,(m)}_{ij}$ are  straightforward 
extensions of the parameters  introduced in \equ{nielsenren}.
Following  the general theorem
of renormalization theory known as Quantum Action Principle (QAP) \cite{QAP},
the terms breaking  the Nielsen identity  at order $n$
 are local polynomial of the fields and we have 
\be\label{app_2.3}
\left[ {\cal S} \left( \Gamma \right) +  \sum_i 
\chi_i \left( \partial_{\xi_i}\Gamma + \Delta_{\chi_i} \cdot \Gamma \right)
\right]^{(n)} = \Delta^{(n)}_0 +  \sum_{i}\chi_i  \Delta^{(n)}_{\chi_i} +
{\cal O}(\chi_i \chi_j),
\ee
where 
the new terms ${\Delta}_{0}^{(n)} + \sum_i \chi_i \Delta^{(n)}_{\chi_i}$ 
are local operators.
We do not consider here the terms ${\cal O}(\chi_j \chi_j)$ as they will 
not enter our forthcoming discussion.
Now we can use the nilpotency of the operator ${\cal S}_{\Gamma_0} +
\sum_{i} \chi_i \partial_{\xi_i}$ to establish the following consistency
conditions for the breaking terms of \equ{app_2.3}:
\bea\label{app_2.4}
 {\cal S}_{\Gamma_0} \Delta^{(n)}_0 = 0,~~~~~~~
 \partial_{\xi_i} \Delta^{(n)}_0 - {\cal S}_{\Gamma_0}
\Delta^{(n)}_{\chi_i} = 0.
\eea
In the absence of anomalies
the first equation can be integrated obtaining the general solution
\cite{kraus,BFM}
\be\label{app_2.5}
\Delta^{(n)}_0 = - {\cal S}_{\Gamma_0} \Gamma^{(n)}_{CT}
\ee
where $\Gamma^{(n)}_{CT}$ are local non-invariant counterterms. These
counterterms are needed to restore the symmetries 
(in our case the STI) to the order $n$ and
are computed by standard techniques of algebraic
renormalization \cite{ghs,sorella}. The removal of the breaking
terms $\Delta^{(n)}_0 $ by means of the counterterms $\Gamma^{(n)}_{CT}$
is essential in order to recover the unitarity of the theory and the physical
interpretation of the S-matrix amplitudes. 
For what concerns the other breaking terms,
namely $ \Delta^{(n)}_{\chi_i}$, they do not play the   essential r\^ole
of the previous ones, but contain the information on the gauge dependence of 
$\Gamma^{(n)}_{CT}$.

The  new functional given by $\g^{(n)}+\g_{CT}^{(n)}$ 
satisfies the STI identity at order $n$. %
On this basis we can study the gauge parameter dependence 
of the Green functions according to the Nielsen identities.
Combining  the second of \eqs{app_2.4}  with \equ{app_2.5} we 
obtain
\be\label{app_2.6}
\partial_{\xi_i} {\cal S}_{\Gamma_0} \Gamma^{(n)}_{CT} + {\cal
S}_{\Gamma_0} \Delta^{(n)}_{\chi_i} = {\cal S}_{\Gamma_0}
\left[\partial_{\xi_i} \Gamma^{(n)}_{C.T.} + \Delta^{(n)}_{\chi_i}
\right] =  0,
\ee
where we have also  used 
$\left[\partial_{\xi_i},  {\cal S}_{\Gamma_0} \right] = 0$
\footnote{In the framework of \efe{kraus} the situation is more complicate as 
the operator ${\cal S}_{\Gamma_0}$ does not commute explicitly with the
derivative wrt the gauge parameters.
}.
Finally, the last equation can be solved  using 
the cohomological methods outlined in Sec.\,2,
\be\label{app_2.7}
\Gamma^{(n)}_{CT} + \Delta^{(n)}_{\chi_i} =
X^{(n)} +  {\cal S}_{\Gamma_0}  Y^{(n)} .
\ee
As discussed in the text, the terms in  $X^{(n)}$ belong to
the cohomology and  represent the gauge parameter dependence of the
{\it physical} parameters. On the other hand, the terms in ${\cal S}_{\Gamma_0}
 Y^{(n)} $ are cohomologically trivial and contribute only to the
{\it unphysical} parameters such as the renormalization of the fields, 
of the gauge  fixing parameters etc. 
Therefore  the insertion of the non-invariant counterterms at 
order $n$ does not affect  ${\beta}_j^{\xi_i,(n)},
{\gamma}_\varphi^{\xi_i,(n)},{\delta}^{\xi_i,(n)}_t$ at the
same order and does not spoil the simple physical interpretation
we have given them in the text.

In summary, we have explicitly seen how the structure of \equ{nielsenren}
is preserved at all orders. When the renormalization conditions are 
chosen according to the scheme presented in Sec.\,\ref{sec2} and all the steps
are properly performed, 
the result of the whole renormalization program are Green functions which
at each order $n$ are finite, satisfy the symmetry properties of the model and
provide S-matrix elements which are bound to be gauge-parameter independent.
\end{appendletterB}

\section*{C. Gauge dependence of the fermionic self-energies}
\begin{appendletterC}
In this appendix we present the explicit gauge-parameter dependence of the
one-loop fermionic unrenormalized self-energies in the SM. 
We consider  the most general case of mixing 
and define the fermionic self-energy $\Sigma_{ij}$ as $+i$ times the standard 
Feynman amplitude for the transition $j\to i$ and extract a factor 
$g^2$. The expressions in the 't Hooft-Feynman gauge ($\xi_i=1$)
can be found, for example, in \efe{denner}.
At the one-loop level, instead of \equ{f1},  we can use the  decomposition
\bea
\Sigma_{ij}(p) = 
\Sigma^{\smallL}_{ij}(p^2)  \not\!p \,P_{\smallL}+ 
\Sigma^{\smallR}_{ij} (p^2) \not\!p \,P_{\smallR} +
\Sigma^{\smallS}_{ij}(p^2) \left( m_i P_{\smallL}+m_j P_{\smallR} \right) \non
\label{B1}.
\eea
The individual components of the self-energies are then given in an arbitrary 
gauge by (similar formulae are also in \cite{fantina}) 
\bea
\Sigma^{\smallS}_{ij}&=& \Sigma^{\smallS}_{ij}|_{\xi=1}
+(\xi_\gamma-1) \,\delta_{ij} \,s_\smallw^2 \,Q_i^2 \,b_{\gamma i}
+(\xi_\smallw-1) \sum_k \lambda^{ij}_k\, \frac{m_k^2}{2} \,c_{\smallw k}
\non\\
&+& (\xi_\smallz-1) \,\frac{\delta_{ij}}{c_\smallw^2}
\left[\ell_i \,r_i \,b_{\smallz i} + \left(\ell_i \,r_i\, \xi_\smallz \,\mz^2 
+\frac{m_i^2}{4}\right) c_{\smallz i}
\right] \label{sigmas}\\ \non\\
%%%%%%%%% sigma L %%%%%%%%%%%%
\Sigma^{\smallL}_{ij}&=& \Sigma^{\smallL}_{ij}|_{\xi=1}\non\\
&+&(\xi_\gamma-1) \,\delta_{ij}\,\frac{s_\smallw^2}{2} \,Q_i^2 
\left[ p^2 (1-x_i)^2 c_{\gamma i} -(1-x_i) \alpha_\gamma -(1+x_i)b_{\gamma
i}\right] 
 \non\\
&+&(\xi_\smallw-1) \sum_k \frac{\lambda^{ij}_k}{4} \left[
 p^2 (1-3x_k) c_{\smallw k}- b_{\smallw k} - \xi_\smallw\mw^2 c_{\smallw k}
-\alpha_\smallw \right]\non\\
&+& (\xi_\smallz-1) \,\frac{\delta_{ij}}{2c_\smallw^2}\left\{
p^2\,c_{\smallz i} \left[\ell_i^2 (1-x_i)^2 -\frac{x_i}{4} (1+x_i)\right]
-\left(\ell_i^2 (1-x_i) +\frac{x_i}{4}\right) \alpha_\smallz \right.\non\\
&-&\left.\left[ \ell_i^2 (1+x_i) - \frac{x_i}{4}\right] 
\left(b_{\smallz i} + \xi_\smallz \,\mz^2 \,c_{\smallz i}\right)
\right\}\label{sigmal}\\ \non\\
%%%%%%%%%%%% sigmaR %%%%%%%%%%%%%%%
\Sigma^{\smallR}_{ij}&=& \Sigma^{\smallR}_{ij}|_{\xi=1}\non\\
&+&(\xi_\gamma-1) \,\delta_{ij}\,\frac{s_\smallw^2}{2} \,Q_i^2 
\left[ p^2 (1-x_i)^2 c_{\gamma i} -(1-x_i) \alpha_\gamma -(1+x_i)b_{\gamma
i}\right] 
 \non\\
&-&(\xi_\smallw-1) \sum_k \lambda^{ij}_k \,\frac{m_i m_j}{4p^2}
\left[\alpha_\smallw -b_{\smallw k} + 
(m_k^2 +p^2 -\xi_\smallw\mw^2) c_{\smallw k}
 \right]\non\\
&+& (\xi_\smallz-1) \,\frac{\delta_{ij}}{2c_\smallw^2}\left\{
p^2 \,c_{\smallz i}  \left[r_i^2 (1-x_i)^2 -\frac{x_i}{4} (1+x_i)\right]
-\left(r_i^2 (1-x_i) +\frac{x_i}{4}\right) \alpha_\smallz \right.\non\\
&-&\left.\left[ r_i^2 (1+x_i) - \frac{x_i}{4}\right] 
\left(b_{\smallz i} + \xi_\smallz \,\mz^2 \,c_{\smallz i}\right)
\right\}\label{sigmar}
\eea
where we have used the following notation for the $n$-dimensional integrals 
($i,j=\gamma,Z^0,W,f$)
 \bea
\alpha_i&=& i  \mu^{4-n} \int \frac{d^n  k}{(2\pi)^n }
\frac1{[k^2 -m_i^2][k^2-\xi_i \,m_i^2]}\non\\
b_{ij}&=& i \mu^{4-n}\int \frac{d^n  k}{(2\pi)^n }
\frac1{[k^2 -m_i^2][(k+p)^2-m_j^2]}\non\\
c_{ij} &=&  i  \mu^{4-n} \int \frac{d^n  k}{(2\pi)^n }
\frac1{[k^2 -m_i^2][k^2 -\xi_i m_i^2][(k+p)^2-m_j^2]}.
\eea
We  have also used $x_i=m_i^2/p^2$, while  
$\ell_i=I_i^3 -Q_i s_\smallw^2$ and $r_i=-Q_i s_\smallw^2$ 
are the left and right-handed couplings of the fermion flavor $i$ and $Q_i$ 
and $I_i^3=\pm\frac1{2}$ its electric charge and isospin.
In the case of quarks, the mixing matrix factor $\lambda_k^{ij}$  
equals $V_{ik} V_{jk}^*$, where $V$ is the CKM matrix, if $i,j$ ($k$) are up
(down) quarks and  $\lambda_k^{ij}=V_{ki}^* V_{kj}$ 
if the opposite is true. For leptons with
massless neutrinos $\lambda_k^{ij}=\delta_{ij}\,\delta_{k\nu_i}$ or 
$\delta_{ij}\,\delta_{kl_i}$, 
i.e.\ there is no mixing.
The gluon exchange diagrams can be obtained from the photonic ones 
setting $Q_i=1$ and multiplying by the color factor $C_F$.
Notice that $\alpha_\gamma$ and $c_{\gamma i}$ are
infrared divergent and an infrared  regulator (like a photon mass) 
should be introduced. Of course, the infrared divergences cancel out in 
 Eqs.(\ref{sigmal}-\ref{sigmar}). It is straightforward to verify 
\cite{hempf}   that in the diagonal case the mass counterterm, 
$\delta m_i/m_i=
\Sigma^\smallS_{ii}(m_i^2)+ \frac1{2} \Sigma^\smallL_{ii}(m_i^2)
+ \frac1{2} \Sigma^\smallR_{ii}(m_i^2) + T_i$,
where $T_i$ is the tadpole contribution, 
is independent of the gauge parameters. From the  off-diagonal parts of 
 Eqs.(\ref{sigmas}-\ref{sigmar}) it is easy  to derive some of the results 
of \efe{ckm} on the gauge dependence of the CKM counterterm.
\end{appendletterC}
%%%%%%%%%%%%%%%%%%%%%%%%%%%%%%% titlepage %%%%%%%%%%%%%%%%%%%%%%%%%%%%%%%%%%%%

\end{document}